\documentclass[11pt]{article}
\pdfoutput=1
\usepackage{jheppub}
\usepackage[T1]{fontenc}
\usepackage[utf8]{inputenc}
\usepackage{cancel,tensor,enumitem,fix-cm}
\usepackage{graphicx}
\usepackage[english]{babel}
\usepackage{amsmath,amssymb}
\usepackage{mathabx}
\usepackage{textcomp}
\usepackage{multirow}
\usepackage{booktabs}
\usepackage{tikz}
\usepackage{overpic}
\usepackage{slashed}

\newcommand{\bea}{\begin{eqnarray}}
\newcommand{\eea}{\end{eqnarray}}
\newcommand{\be}{\begin{equation}}
\newcommand{\ee}{\end{equation}}

\def\({\left(}
\def\){\right)}

\def \l {\lambda}
\def \m {\mu}
\def \n {\nu}

\def \t {\tau}

\def \D {\Delta}

\def\nn {\nonumber}

\subheader{\begin{flushright}
\texttt{MIT-CTP/5255, BRX-TH-6670, APCTP Pre2020-031}
\end{flushright}}

\title{On the interplay between magnetic field and anisotropy in holographic QCD}

\author[a]{Umut G\"ursoy,}
\author[b,c,d]{Matti J\"arvinen,}
\author[a,e]{Govert Nijs}
\author[f,g]{and Juan F. Pedraza}
\affiliation[a]{Institute for Theoretical Physics and Center for Extreme Matter and Emergent Phenomena,\\ Utrecht University, 3584 CC Utrecht, The Netherlands}
\affiliation[b]{Raymond and Beverly Sackler School of Physics and Astronomy, Tel-Aviv University,\\ Tel-Aviv 69978, Israel}
\affiliation[c]{Asia Pacific Center for Theoretical Physics, Pohang, 37673, Korea}
\affiliation[d]{Department of Physics, POSTECH, Pohang, 37673, Korea}
\affiliation[e]{Center for Theoretical Physics, Massachusetts Institute of Technology, Cambridge, MA 02139, USA}
\affiliation[f]{Department of Physics and Astronomy, University College London, London WC1E 6BT, UK}
\affiliation[g]{Martin Fisher School of Physics, Brandeis University, Waltham MA 02453, USA}
\emailAdd{u.gursoy@uu.nl}
\emailAdd{matti.jarvinen@apctp.org}
\emailAdd{govert@mit.edu}
\emailAdd{j.pedraza@ucl.ac.uk}

\abstract{We investigate the combined effects of anisotropy and a magnetic field in strongly interacting gauge theories by the gauge/gravity correspondence. Our main motivation is the quark-gluon plasma produced in off-central heavy-ion collisions which exhibits large anisotropy in pressure gradients as well as large external magnetic fields. We explore two different configurations, with the anisotropy either parallel or perpendicular to the magnetic field, focusing on the competition and interplay between the two. A detailed study of the RG flow in the ground state reveals a rich structure where depending on which of the two, anisotropy or magnetic field, is stronger, intermediate geometries with approximate AdS$_4\times \mathbb{R}$ and AdS$_3\times \mathbb{R}^2$ factors arise. This competition is also manifest in the phase structure at finite temperature, specifically in the dependence of the chiral transition temperature on anisotropy and magnetic field, from which we infer the presence of inverse magnetic and anisotropic catalyses of the chiral condensate. Finally, we consider other salient observables in the theory, including the quark-antiquark potential, shear viscosity, entanglement entropy and the butterfly velocity. We demonstrate that they serve as good probes of the theory, in particular, distinguishing between the effects of the magnetic field and anisotropy in the ground and plasma states. We also find that the butterfly velocity, which codifies how fast information propagates in the plasma, exhibits a rich structure as a function of temperature, anisotropy and magnetic field, exceeding the conformal value in certain regimes.}

\begin{document}
\maketitle
\flushbottom

\section{Introduction}
\begin{figure}[t!]
\centering
\includegraphics[width=0.5\textwidth]{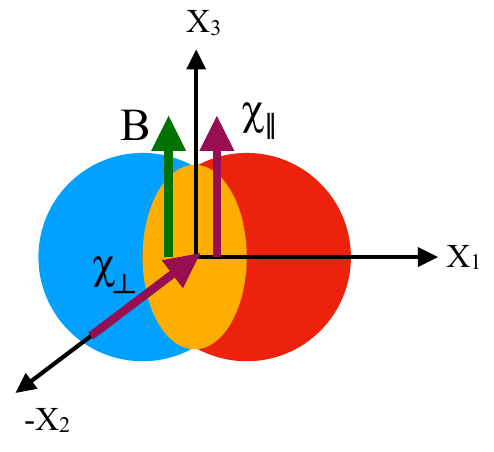}
\caption{Schematic picture of an off-central heavy-ion collision. The beam direction is denoted by $x_2$\@. Red and blue circles represent the colliding gold/lead nuclei moving in the $+x_2$ and $-x_2$ directions respectively. They produce an almond shaped plasma here shown in orange. The amount of anisotropy in the stress-energy tensor is represented by the $\chi$ vectors, drawn in purple. $\chi_\parallel$ controls the amount of anisotropy in the interaction plane $(x_1,x_3)$ and $\chi_\perp$, which is present also in central collisions, distinguishes plasma pressure in the $x_2$ direction --- in which it expands faster --- and the interaction plane. A magnetic field in the $x_3$ direction (drawn in green) is also produced (only in off-central collisions) mostly by the nucleons that do not participate in the formation of the plasma but fly away (here represented by the non-overlapping  portions of the blue and red circles).\label{fig:HIC}}
\end{figure}

Studies of the quark-gluon plasma produced in heavy-ion collisions at RHIC and LHC have revealed an accepted description of this plasma as a strongly interacting relativistic fluid, see \cite{Busza:2018rrf} for a recent review. In particular, anisotropies in soft particle distributions detected in off-central collisions played a central role in establishing this description early on \cite{Heinz:2013th,Teaney:2000cw,Romatschke:2007mq}. According to the picture that emerged from these studies, the initial almond shape of the plasma created in off-central collisions leads to different pressure gradients building up in the two directions --- which we denote by $x_1 \equiv x$ (short axis of the almond) and $x_3 \equiv z$ (long axis of the almond), see fig.~\ref{fig:HIC} ---  that are transverse to the beam direction --- which we denote by $x_2 \equiv y$, and these different pressure gradients in turn yield different multiplicities of hadrons detected in these directions. This anisotropy can be quantified by introducing the so-called elliptic flow coefficient $v_2\{2\}$. Matching of this parameter in the experiment to viscous hydrodynamics simulations results in a relativistic fluid with extremely small shear viscosity to entropy ratio \cite{Novak:2013bqa,Pratt:2015zsa,Sangaline:2015isa,Bernhard:2016tnd,Bernhard:2019bmu,Devetak:2019lsk,Auvinen:2020mpc,Moreland:2018gsh,Everett:2020yty,Nijs:2020ors,Nijs:2020roc,Everett:2020xug} consistent with the universal value that follows from the AdS/CFT correspondence \cite{Policastro:2001yc}.

In this accepted hydrodynamic picture, the plasma also expands mainly in the $x_2$ direction, see fig.~\ref{fig:HIC}. This brings in another anisotropy in the system, that is present also in central collisions, which stems from the pressure in the $x_2$ direction being different from the transverse pressures on the interaction plane). In an off-central collision, in general, the spatial components of the stress-energy tensor in all directions, hence, are different from each other.

This picture is further complicated by the strong magnetic fields produced in off-central collisions mainly sourced by the positively charged ``spectator'' ions that fly off without participating in plasma formation (fig.~\ref{fig:HIC}) \cite{Tuchin:2013ie,Skokov:2009qp}. Thus, this anisotropic plasma is also produced in the presence of an external magnetic field in the $x_3$ direction in our conventions. Magnetic fields lead to interesting effects in QCD\@. First, there is anomaly induced transport of electric and chiral charge \cite{Fukushima:2008xe} sourced by magnetic fields (and also vorticity). These chiral magnetic and vortical effects are proposed as possible solutions to the strong CP problem \cite{Kharzeev:2015znc,Shi:2017cpu}. Having already been observed in Dirac semimetals \cite{Li:2014bha}, the Chiral Magnetic Effect (CME) is yet to be seen in heavy ion collisions. The main difficulty in its experimental detection precisely lies in disentangling the similar effects that result from the anisotropies in the transverse $x_1x_3$ plane  that we mentioned above. Therefore, a distinguishing signature of dynamics due to anisotropies and magnetic fields is quite important in this quest, and it is one of the motivations behind our work.

Second, there are the so-called magnetic catalysis (MC) and the inverse magnetic catalysis (IMC) effects, which are respectively the strengthening and weakening of the chiral condensate in the hadronic phase by a magnetic field. Which one is realized depends on the magnitude of the magnetic field and the temperature. While the physical mechanism behind the former is relatively well understood \cite{Miransky:2015ava}, the IMC, which was first observed on the lattice \cite{Bali:2012zg} remains an open question. In a recent work \cite{Gursoy:2018ydr}, based on the holographic model we discuss below and following the earlier insights of \cite{Giataganas:2017koz}, we observed that a similar effect results directly from anisotropy in a holographic QFT\@. Thus, once again, it becomes crucial to distinguish  the signatures of anisotropy from that of the magnetic field.

In this paper, we take on this task and employ a holographic effective theory for large-$N$ QCD to explore the interplay between anisotropy and magnetic field. As mentioned above both are naturally present in heavy ion collisions. In particular, anisotropy in the pressure gradients is realized parallel to a magnetic field in the transverse plane in the off-central collisions, while an additional anisotropy which results from expansion of the plasma in the beam direction is realized perpendicular to the magnetic field. Two extreme instances of anisotropy without magnetic field are i) anisotropy in the beam direction in the central collisions and ii) anisotropy both in the transverse and the beam directions in the central collisions of Uranium ions which are naturally almond shaped. We will investigate in this paper the intricate interplay in these settings in an idealized setup where we treat the plasma as completely equilibrated and with infinite extent.

The holographic setup that we consider in this paper is based on an extension of the bottom-up model called improved holographic QCD (ihQCD) \cite{Gursoy:2007cb,Gursoy:2007er} that includes fermion flavors, known as V-QCD~\cite{Jarvinen:2011qe}. It provides a detailed effective description of both the confined and the deconfined phases of large-$N$ confining gauge theories realizing the salient qualitative features of QCD such as linear confinement of quarks, running of the gauge coupling with IR slavery, confinement-deconfinement and chiral phase transitions. Moreover, it involves a sufficient number of parameters that can be fitted to the available lattice data to provide a quantitative description of the theory. As we argue in this paper, a further extension of this model that includes a magnetic field and anisotropy will have predictive value for the quark-gluon plasma that results in off-central heavy ion collisions and will lead to interesting qualitative insights and understanding of the underlying strong dynamics.

The magnetic field $B$ is introduced in the model in the spatial components of the diagonal $U(1)_{L+R}$ gauge field on the space-filling flavor branes. We take $B$ constant and in the $x_3$ direction throughout this paper. We further introduce anisotropy through a bulk scalar field, `axion' $\chi$, whose profile is chosen to be linear either in the $x_3$, $\chi_\parallel = a x_3$ or in the $x_2$, $\chi_\perp = a x_2$ (see fig.~\ref{fig:HIC}) direction, where $a$ is a parameter that measures the magnitude of anisotropy. This axion is dual to the topological charge operator in the dual gauge theory, hence these choices correspond a space dependent $\theta$ angle that breaks the rotation symmetry in the boundary theory. One can rotate $\theta$ away by a chiral transformation, producing a constant external axial gauge field which couples to the fermions in the theory similar to the magnetic field \cite{Gursoy:2018ydr} and hence competes with it. We probe this competition between anisotropy and magnetic field by varying the value of $a$ and $B$ and studying the resulting effects on physical observables, i.e.~the chiral transition temperature, quark-antiquark potential, string tension, entanglement entropy and the butterfly velocity. Definitions of these quantities and their holographic representations are detailed in the respective sections.

The paper is organized as follows. In the next section we present the holographic model in detail. In section \ref{app:asymptotics} we study the RG flow in the theory. The competition between anisotropy and magnetic field manifests itself in the holographic RG flow by approximate AdS$_4$ and AdS$_3$ bulk regions which respectively arise when anisotropy dominates over the magnetic field, and vice versa. The far IR regime of our solutions is dominated by an AdS$_4$ (up to logarithmic corrections in the holographic coordinate) as the flavor branes decouple in the IR, hence the magnetic field gets screened but anisotropy prevails. In section \ref{sec::thermo} we study how the phase structure of the theory depends on $a$ and $B$. In particular we study how the chiral transition temperature depends on these parameters and show that the theory exhibits both inverse magnetic and anisotropic catalyses at low temperatures. Section \ref{sec:otherObs} is devoted to study of other observables, i.e.~the quark-anti-quark potential, string tension, shear viscosities, entanglement entropy and the butterfly velocity. These observables turn out to be useful probes of phenomena associated to anisotropy and magnetism, and are used to study the competition between the two.  We show that linear confinement in the ground state in the absence of anisotropy and magnetic field is broken in the presence of the former but not the latter. Shear viscosities on the three separate planes monotonically decrease from UV to IR from the universal holographic value corresponding to the isotropic theory toward substantially smaller values. We observe that entanglement entropy, in addition to the quark-antiquark potential, can be used as a good order parameter for confinement/deconfinement. Finally the butterfly velocity, which codifies how fast information propagates in the plasma, exhibits a rich structure as a function of temperature, anisotropy and magnetic field, exceeding the conformal value in certain regimes. We summarize and discuss our results in section \ref{sec::discuss}, also providing an outlook. Details of our calculations are given in two appendices.

\section{Holographic setup}\label{sec:setup}

\subsection{Gravitational action}

The holographic model we consider includes the backreaction of quark flavors to glue dynamics. The gravity action is composed of two sectors:
\be
S=S_g+S_f\,,
\label{action}
\ee
The glue sector $S_g$ is formally given by the improved holographic QCD (ihQCD) model \cite{Gursoy:2007cb,Gursoy:2007er} which will be deformed here by an extra dimension-$4$ operator $\mathcal{O}_\chi$ that is responsible for the breaking of the spatial $SO(3)$ symmetry \cite{Gursoy:2018ydr}. The flavor sector $S_f$ is based on a generalized tachyonic Dirac-Born-Infeld action that arises from a pair of space filling $D4-\overline{D4}$ branes~\cite{Bigazzi:2005md,Casero:2007ae}. This sector controls the physics of chiral symmetry breaking and also allows for the introduction of a background magnetic field. The two sectors fully backreact in the Veneziano limit where $N_f \to \infty$ and $N_c \to \infty$ and $N_f/N_c \equiv x_f$ is fixed. This backreacted model is known as V-QCD~\cite{Jarvinen:2011qe}.

The ihQCD action for the glue sector is given explicitly as
\be
S_g= M^3 N_c^2 \int d^5x \ \sqrt{-g}\left(R-{4\over3}{
(\partial\lambda)^2\over\lambda^2}+V_g(\lambda)-\frac{1}{2}Z(\lambda)(\partial \chi)^2\right) \, ,
\label{actg}
\ee
where $M$ is an effective Planck mass that is fixed by the UV asymptotics of the free energy. In this sector we have three bulk fields that are dual to relevant or marginal operators that dictate the IR dynamics: the stress-energy tensor $T_{\mu\nu}$, dual to the metric $g_{\mu\nu}$, the glueball operator $\mathcal{O_\lambda}\sim\text{Tr}\, F^2$ dual to the dilaton $\l$ and a pseudo-scalar operator $\mathcal{O}_\chi\sim\text{Tr}\,F\wedge F$ dual to the axion field $\chi$. As mentioned above, the latter operator is introduced in order to break rotational symmetry, as in \cite{Gursoy:2018ydr} (see also \cite{Mateos:2011ix,Mateos:2011tv,Giataganas:2017koz} where this mechanism was first implemented in other holographic models). This breaking is achieved by adding a source that depends on one of the field theory directions explicitly breaking the $SO(3)\to SO(2)$ spatial symmetry. We note that a proper implementation of the QCD axial anomaly would require coupling of axion to the flavor sector, which enters at leading order in the Veneziano limit~\cite{Arean:2013tja,Arean:2016hcs}. However, since we only introduce the axion as a means to break isotropy, such couplings will not be essential for our purposes and we choose to omit them for simplicity.

The action of the flavor sector reads
\begin{align}
\label{actf}
S_f =-x_f\, M^3 N_c^2 \int d^5x V_f(\l,\t) \sqrt{- \mathrm{det}\left(g_{\m\n} + \kappa(\l)\, \partial_{\m} \t \,\partial_{\n} \t+ w(\l)\, V_{\m\n}\right) } \,.
\end{align}
where we only wrote down explicitly the terms in the DBI sector which are relevant for the analysis in this article (see~\cite{Casero:2007ae,Arean:2013tja} for a complete description). This sector includes two additional fields, the tachyon $\t$, dual to the quark bilinear operator $\bar{q} q$, and the field strength $V_{\mu\nu}=\partial_{\mu} V_{\nu}-\partial_{\nu} V_{\mu}$. The presence of the tachyon is crucial to our model since it serves as an order parameter for chiral symmetry breaking \cite{Casero:2007ae}, whereas the field strength will be used to introduce a boundary magnetic field $B$.

Considering the above, we take the following ansatz for the metric and bulk fields:
\begin{align}\label{ansatz}
ds^2&=e^{2A(r)}\left[-f(r) dt^2+dx_1^2+e^{2U(r)}dx_2^2  + e^{2W(r)} dx_3^2+\frac{dr^2}{f(r)}\right]\,,\\
\lambda&=\lambda(r)\,,\qquad\tau=\tau(r)\,.
\end{align}
The gauge field is chosen such that it generates an external magnetic field in the $x_3$ direction at the boundary:
\begin{align}
\qquad V_{\mu}= \left(0, -x_2 B/2, x_1 B/2,0,0 \right)\,.
\end{align}
Notice that we did not introduce flavor dependence in the action~\eqref{actf}: we only have a single type of quark flavors and all quarks have the same mass. Therefore the magnetic field couples in the same way to all flavors, i.e., it couples to the baryon number rather than the electric charge.
We stress that in our construction the magnetic field $B$ enters through the flavor sector, in contrast to many other simpler models, e.g.~\cite{DHoker:2012rlj,Rougemont:2014efa,Rougemont:2015oea,Dudal:2016joz}, 
where $B$ is introduced via a Maxwell term which couples directly to the geometry.
For the field $\chi$ we consider two qualitatively different options:\footnote{The most general case would be $\chi = a\sin(\theta)x_2 + a\cos(\theta)x_3 \equiv a_2x_2 + a_3x_3$. However, this generates an off-diagonal term in the Einstein equations, requiring a non-diagonal metric ansatz. We will therefore focus on the two cases above, i.e., parallel and perpendicular to the magnetic field, for simplicity.}
\be
\chi = \chi_\parallel \equiv a\, x_3\qquad\text{or}\qquad \chi=\chi_\perp \equiv a\, x_2\,.
\ee
This ansatz automatically satisfies the equations of motion for the axion $\chi$ and the gauge field $V_\mu$, while introducing anisotropy in the $x_3$ direction (in the parallel case $\chi=\chi_\parallel$) or both in the $x_2$ and $x_3$ directions (in the perpendicular case $\chi=\chi_\perp$). The amount of anisotropy is controlled by the parameters $a$ and $B$. The metric functions $W$ and $U$, on the other hand, codify the dependence of the anisotropy along the RG flow. For a detailed account of the non-trivial equations of motion see Appendix \ref{App:eqns}.

\subsection{Choice of potentials} \label{sec:pots}

Constraints to the potential functions and couplings in the action from various sources have been discussed in detail in earlier literature~\cite{Gursoy:2007cb,Gursoy:2007er,Gursoy:2009jd,Gursoy:2012bt,Jarvinen:2011qe,Alho:2012mh,Arean:2013tja,Jarvinen:2015ofa,Arean:2016hcs,Jokela:2018ers}. In the current study, the  coupling $Z(\lambda)$ between the dilaton and the axion is taken from \cite{Gursoy:2012bt,Drwenski:2015sha} while the other potentials $V_g(\lambda)$, $V_f(\lambda,\tau)$, $\kappa(\lambda)$ and $w(\lambda)$
are taken from \cite{Alho:2012mh,Alho:2013hsa,Gursoy:2016ofp}. We work at $x_f=N_f/N_c=1$ in this article, corresponding roughly to $N_f=3$ dynamical light quarks at $N_c=3$. The potentials are given explicitly in Appendix~\ref{app:potentials}; we comment here on their general properties.

We require generic agreement with qualitative features of QCD such as asymptotic freedom, confinement, and linear meson trajectories. That is,
at the boundary (UV) we require that $\lambda \to 0$ (asymptotic freedom) and $\tau \to 0$ for all solutions, which is obtained simply if near $\lambda = 0$ all potentials take constant values up to $\mathcal{O}(\lambda)$ corrections~\cite{Gursoy:2007cb,Gursoy:2007er,Jarvinen:2011qe,Arean:2013tja}. The asymptotic values are among other things fixed by the canonical dimensions of the dual operators.

The UV behavior acts as a boundary condition for the more interesting and rich strongly coupled IR dynamics. The IR dynamics in the model is to a large extent determined by the asymptotic behavior of the potentials as $\lambda \to \infty$, which is in turn constrained by confinement, linear radial glueball and meson trajectories, regular IR behavior of the bulk solutions, mass gap of the meson spectrum at large quark mass, and regularity of the solutions at finite $\theta$-angle~\cite{Gursoy:2007cb,Gursoy:2007er,Jarvinen:2011qe,Arean:2013tja,Jarvinen:2015ofa,Arean:2016hcs}. Interestingly, these considerations seem to point towards asymptotics which match with the expectations from string theory (certain power laws of $\lambda$) up to logarithmic corrections~\cite{Gursoy:2007er,Arean:2013tja,Arean:2016hcs}.

With the asymptotics of the potentials determined both as $\lambda \to 0$ and as $\lambda \to \infty$, it remains to fix their behavior in the middle. In the end this needs to be done by comparing to QCD data from experiments or from lattice, as in~\cite{Gursoy:2009jd,Panero:2009tv,Jokela:2018ers}. However, in a model with such a large number of parameters as the current model, the potentials need to be regular monotonic functions in practice in order to avoid unphysical behavior. Therefore there is relatively little freedom left to fit in the middle. The main parameter is the rough value of $\lambda$ where one moves from the UV region to the IR region for each potential. For magnetic phenomena, the coupling of the gauge field, $w(\lambda)$, is particularly important. For this potential, the location of the UV to IR ``transition'' is controlled through the parameter $c$ (which does not affect the asymptotics), defined through
\be\label{def:c}
 w(\lambda) = \kappa(c \lambda) \,.
\ee

This Ansatz was studied in~\cite{Gursoy:2016ofp}, and it was found that the parameter $c$ affects the strength of the inverse magnetic catalysis, i.e., how the chiral condensate is suppressed with increasing magnetic field at temperatures around the chiral and deconfinement transitions of the model. Agreement with lattice data was found at relatively small values of $c$ such as $c=0.4$ and $c=0.25$. The model was independently fitted to lattice data for QCD thermodynamics in~\cite{Jokela:2018ers} using a slightly different Ansatz from the current article, which anyhow implements the same asymptotic behavior for all functions. It was found that the baryon number susceptibility
\be
\chi_B = \left.\frac{d^2P}{d\mu^2}\right|_{\mu=0}
\ee
is sensitive to the choice of $w$. Interestingly, a good fit to the lattice data required\footnote{The parameter which maps to $c$ of this article was denoted as $w_s$ in~\cite{Jokela:2018ers}.} that $c \approx 1/3$, i.e., a value in agreement with the results of~\cite{Gursoy:2016ofp}.
In this article, we will discuss additional observables which agree with QCD data for similar values of $c$.

\section{RG flow: from the UV to the IR} \label{app:asymptotics}

In this section, we discuss the basic structure of the (vacuum, zero temperature) geometry for various values of $a$ and $B$. We analyze the asymptotic UV (small coupling) geometry, the asymptotic IR (large coupling) geometry, and the RG flow in between. As we shall see, for large values of $a$ and/or $B$ there are also interesting intermediate scaling regions in the flow.

\subsection{UV asymptotics}

We start by checking the UV asymptotics. As it turns out, it is only affected by the inclusion of $a$ and/or $B$ in highly subleading terms. That is, the ``source'' terms in the asymptotic behavior of the coupling and the metric are independent of both $a$ and $B$.

The UV structure is set up~\cite{Gursoy:2007cb,Gursoy:2007er,Jarvinen:2011qe} such that
\begin{align} \label{UVexpsapp}
  V_1 \l(r)&=-\frac{8}{9 \log(r \Lambda)} + \frac{
   \log\left[-\log(r \Lambda)\right] \left[\frac{46}{81} - \frac{128 V_2}{81
V_1^2}\right]}{\log[r \Lambda]^2}+{\cal
O}\left(\frac{1}{\log(r\Lambda)^3}\right) \nn \\\nn
A(r) &= -\log\frac{r}{\ell} + \frac{4}{9 \log(r \Lambda)}  \\
&\phantom{=}+ \frac{
  \frac{1}{162} \left[95  - \frac{64 V_2}{V_1^2}\right] +
   \frac{1}{81} \log\left[-\log(r \Lambda)\right] \left[-23 + \frac{64
V_2}{V_1^2}\right]}{
  \log(r \Lambda)^2} +{\cal O}\left(\frac{1}{\log(r\Lambda)^3}\right) \ ,
\end{align}
where the coefficients are defined by
\be \label{Videfapp}
 V_{\rm eff}(\l)=V_g(\l)-x_f V_f(\l,0)=\frac{12}{\ell^2}\left[1 + V_1 \l +V_2
\l^2+\cdots \right] \ .
\ee
The other functions $f$, $\exp (W)$, and $\exp (U)$ tend to constants in the UV up to highly suppressed corrections, and without loss of generality these constants can be set to one.  See Appendix~\ref{app:thermodynamics} for the expansions and the discussion of the normalizable terms in the expansions. In analogy to the dimensional transmutation of QCD, the expansions~\eqref{UVexpsapp} define the characteristic scale $\Lambda$, which we will use to normalize dimensional quantities below.

\subsection{IR asymptotics at $a \ne 0$}

We then discuss the IR behavior of the vacuum (zero temperature) solutions.
The IR asymptotics depends on the choice of potentials, and we only discuss here the results for the potentials used in this article (see Sec.~\ref{sec:pots} and Appendix~\ref{app:potentials}), which were determined by comparing with QCD.
At zero magnetic field the asymptotics has been analyzed in the literature~\cite{Gursoy:2007cb,Gursoy:2007er,Jarvinen:2011qe,Arean:2013tja,Gursoy:2018ydr}.
In particular,
the IR asymptotics of the thermodynamically preferred, chirally broken solution in the presence of a nonzero anisotropic parameter $a$ was studied in detail in~\cite{Gursoy:2018ydr}. The geometry was seen to be asymptotically AdS$_4\times \mathbb{R}$ up to logarithmic corrections. The IR geometry therefore changes drastically as any nonzero $a$ is turned on.\footnote{Notice that at finite charge (baryon number) there is a similar change in the IR geometry, which in that case is AdS$_2 \times \mathbb{R}^3$~\cite{Alho:2013hsa}.} We show here that the same geometry is found also when the magnetic field is nonzero.

We consider the asymptotics of the full system, i.e., $x_f >0$, $a \ne 0$, and possibly $B \ne 0$. We first assume that the tachyon diverges fast enough so that it decouples the flavor from the glue in the IR, which will be verified below by using the final result. In the absence of the flavor sector and for nonzero $a$ the metric is asymptotically AdS$_4\times \mathbb{R}$. For the case $a=a_\parallel$, for example, the geometry is given by~\cite{Gursoy:2018ydr}
\be \label{IRasympt}
 e^{A} \sim \frac{1}{r}e^{-\sqrt{(\log r)/6}-(\log\log r)/8} \ , \quad e^{W+A} = e^{\widetilde W} \sim e^{\sqrt{(2\log r)/3}-(\log\log r)/8} \ , \quad \phi \sim \sqrt{(3\log r)/8} \ ,
\ee
so that $\widetilde W$ and $\phi$ vary much slower than $A$ while $f$ and $U$ approach constants in the IR, up to highly suppressed corrections.
It is useful to write the tachyon equation of motion in a different form than~\eqref{taueom}. Assuming\footnote{See appendix \ref{app:potentials} for the definition of constant $a_0$.} $V_f(\l,\tau) = V_{f0}(\l)e^{-a_0\tau^2}$, it can be rearranged as
\be \label{taueq2}
 \frac{G}{e^{5A+W+U}V_{f0}(\l)Q}\frac{d}{dr}\left[\frac{e^{3 A+W+U} f \tilde{\kappa} (\lambda) V_{f0}(\lambda) Q \dot\tau}{G}\right] = -2 \tau \ ,
\ee
where $\tilde{\kappa} \equiv \kappa /a_0$.
Since $\l$ evolves slowly in the IR and the tachyon diverges, we may approximate $G \simeq e^{-A} \sqrt{\kappa(\l)} \dot\tau$. Neglecting derivatives of $\l$  and $f$ we find that
\be
 \frac{f \tilde{\kappa} (\lambda) \dot\tau}{e^{6A+W+U}Q}\frac{d}{dr}\left[e^{4 A+W+U}   Q \right] \simeq -2 \tau \ .
\ee
For $B \ne 0$, we find $Q \simeq e^{-2A-U} w(\l) |B|$, and when $B=0$ we have $Q=1$ and also $f=1$ since we are studying the zero temperature solution. We further insert the rough approximation $e^{-A} \simeq r/q$ where $q$ is constant, and approximate $W\sim -A$, $U \sim \mathrm{const}$ for $a=a_\parallel$ and $U\sim -A$, $W \sim \mathrm{const}$ for $a=a_\perp$.
We obtain
\begin{align}
 \frac{3 \tilde{\kappa} (\lambda) r \dot\tau }{2q^2} &\simeq \tau \ ,&  &(B=0)& \nonumber\\
 \frac{ f \tilde{\kappa} (\lambda) r \dot\tau }{2q^2}  &\simeq \tau \ ,& &(B\ne0\ , \quad a=a_\parallel)&\\\nonumber
 \frac{ f \tilde{\kappa} (\lambda) r \dot\tau }{q^2}  &\simeq \tau \ ,& &(B\ne0\ , \quad a=a_\perp)\ .&
\end{align}
The solution is therefore
\begin{align}
 \tau &\sim r^{2 q^2/(3 \tilde{\kappa}(\l))}\ ,\quad (B=0) \ ; \qquad\qquad \tau \sim r^{2 q^2/(f \tilde{\kappa}(\l))}\ ,\quad (B\ne 0\ , \quad a=a_\parallel)&\nonumber\\
 \tau &\sim r^{ q^2/(f \tilde{\kappa}(\l))}\ ,\quad (B\ne 0\ , \quad a=a_\perp)\ .&
\end{align}
For the (approximately) AdS$_4$ IR asymptotics, $q^2 V_g \sim \mathrm{const}$~\cite{Gursoy:2018ydr}, when $q$ is promoted into a slowly varying field.
Moreover because $f = \mathrm{const}$ and we have chosen the potentials such that $V_g \tilde{\kappa} \sim \mathrm{const}$ at large $\l$, we find that $q^2/(f \tilde{\kappa}(\l)) \sim \mathrm{const}$. Therefore the tachyon indeed obeys a power law in the IR. This is enough for the tachyon to decouple and for the assumptions we made above to be valid.

\subsection{IR asymptotics at $a=0$ and $B \ne 0$}

For completeness, we also discuss the IR asymptotics of the model in the absence of anisotropy (so that $U=0$). Again first assuming that the flavor sector is asymptotically decoupled in the IR, which we will verify below, the asymptotics of the geometry are given by~\cite{Jarvinen:2011qe}
\be \label{bgstandardIR}
 A = -r^2 + \frac{1}{2} \log r + A_c + \mathcal{O}\left(\frac{1}{r^2}\right) \ ,\qquad \log \lambda = \frac{3 }{2}r^2 + \lambda_c +  \mathcal{O}\left(\frac{1}{r^2}\right)
\ee
where we set the scale of expansions to one. The constants $A_c$ and $\lambda_c$ have complicated expressions in terms of the IR expansions of the potentials which we have omitted for simplicity. The functions $W(r)$ and $f(r)$ tend to constants in the IR up to highly suppressed corrections.

As above, it is then enough to study the asymptotic behavior of the tachyon, determined by~\eqref{taueq2}, and check that the tachyon indeed grows fast enough to decouple the flavor action from the metric. As above, we may approximate $G \simeq e^{-A} \sqrt{\kappa(\l)} \dot\tau$. Then the tachyon equation of motion can be approximated as
\be
 e^{-2A} f \tilde{\kappa} (\lambda) \dot\tau \frac{d}{dr}\log\left[e^{4 A}\sqrt{\tilde{\kappa}(\lambda)} V_{f0}(\lambda)   Q \right] \simeq -2 \tau \ .
\ee
The result for quite generic choices of $\tilde{\kappa}$, $V_{f0}$, and $w$ can be obtained by integrating this equation (see~\cite{Arean:2013tja} for a detailed analysis at zero magnetic field). Here we restrict to the potentials used in this article. For the choice of $w$ in~\eqref{wpotential}, we note that $Q \to 1$ in the IR, so that the dependence on the magnetic field in the asymptotic tachyon equation only appears through the IR value of the function $f$. Inserting the asymptotics for the other potentials and the background from~\eqref{bgstandardIR} we obtain that $\tau \sim \exp(\# r^2)$ with a positive coefficient in the argument of the exponential, which is enough to decouple the flavors in the IR.

\subsection{Intermediate energies at large $B$ \label{sec:intermB}}

It may appear surprising that while in the presence of finite $a$ the IR geometry is strongly modified with respect to the solution at $a =0$, for finite $B$ (and zero $a$) the IR asymptotics are the same as for $B=0$. This happens because the anisotropic term in~\eqref{actg} couples directly to the geometry, but the magnetic field is included as part of the flavor action~\eqref{actf}, which decouples in the IR. Therefore it is natural that the IR geometry is insensitive to the value of the magnetic field.

However, when the magnetic field is large, there is a region in the bulk geometry where the coupling of the magnetic field to the geometry is large but the IR decoupling of the flavors has not yet started. In this region, as we shall argue below, the geometry is roughly AdS$_3\times\mathbb{R}^2$ in analogy to the IR geometry AdS$_4\times\mathbb{R}$ of finite $a$.

The  AdS$_3\times\mathbb{R}^2$ geometry can actually be found as a constant scalar solution to the V-QCD action in a (double) scaling limit. For the tachyon we set $\tau=0$ indicating that the chiral symmetry is fully intact and the decoupling of the flavors is absent. For simplicity we set $a=0$ so that $U=0$.  Because boosts in the direction of the magnetic field are unbroken, and we are looking for a zero temperature solution, we also take $f=e^{2 W}$. Keeping terms up to $\mathcal{O}\left(B^0\right)$ the relevant Einstein equations simplify to
\begin{align}
\frac{1}{2}x |B|  e^{-2 W} V_f(\lambda_c,0)\,w(\lambda_c)-3 \dot A \dot W-3 \dot A^2-2 \dot W^2+3 \ddot A &=0 \,, & \\
\frac{1}{2} e^{2 A-2 W}V_g(\lambda_c) -\frac{1}{2}x |B|  e^{-2 W} V_f(\lambda_c,0)\,w(\lambda_c)-6 \dot A \dot W-6 \dot A^2-\dot W^2 &=0 \,, &
\end{align}
where $\lambda_c$ is the constant value of the dilaton. A precise way of deriving this set of equations is to take the double scaling limit with $x_f \to 0$ and $|B| \to \infty$ with $x_f |B|$ fixed.
There is a unique ``scaling'' solution which satisfies these equations, given by
\be
 A(r) = A_0 \,, \qquad W(r) = \log(r) + W_0
\ee
where the constants are defined in terms of the equations
\be
 e^{2 W_0} = \frac{1}{4}x |B|\,  V_f(\lambda_c,0)\,w(\lambda_c) \,, \qquad e^{2A_0} = \frac{3x |B|\,  V_f(\lambda_c,0)\,w(\lambda_c)}{2 V_g(\lambda_c)} \,.
\ee
The geometry becomes
\be \label{ads3geom}
ds^2 = e^{2 A_0+2W_0} r^2 (-dt^2+dx_3^2)  + e^{2 A_0-2W_0} r^{-2} dr^2 + e^{2A_0}(dx_1^2+dx_2^2)\,,
\ee
which is indeed recognized as AdS$_3\times\mathbb{R}^2$.

For this solution to be exact in the double scaling limit, the dilaton equation of motion imposes the further condition
\be
 3 \frac{d}{d \lambda}\log V_g(\lambda) -2 \frac{d}{d \lambda}\log w(\lambda)-2 \frac{\partial}{\partial \lambda}\log V_f(\lambda,0)\bigg|_{\lambda=\lambda_c} = 0 \,,
\ee
which is analogous to the condition found in the case of anisotropic IR asymptotics for $a \ne 0$ in~\cite{Gursoy:2018ydr} (see Eq.~(3.7) in this reference). The potentials used in this article do not admit a solution to this condition. However, an AdS$_3\times\mathbb{R}^2$ region may still appear as an approximation in the geometries at large $B$.
Indeed, as we discuss below, a clearly recognizable  AdS$_3\times\mathbb{R}^2$ region is found in the numerical solutions of the full RG flows at large values of the magnetic field. In this region, as expected from the constant scalar analysis, the tachyon is small while the dilaton flows much slower than $W$.

\begin{figure}[t!]
\centering
\includegraphics[width=0.7\textwidth]{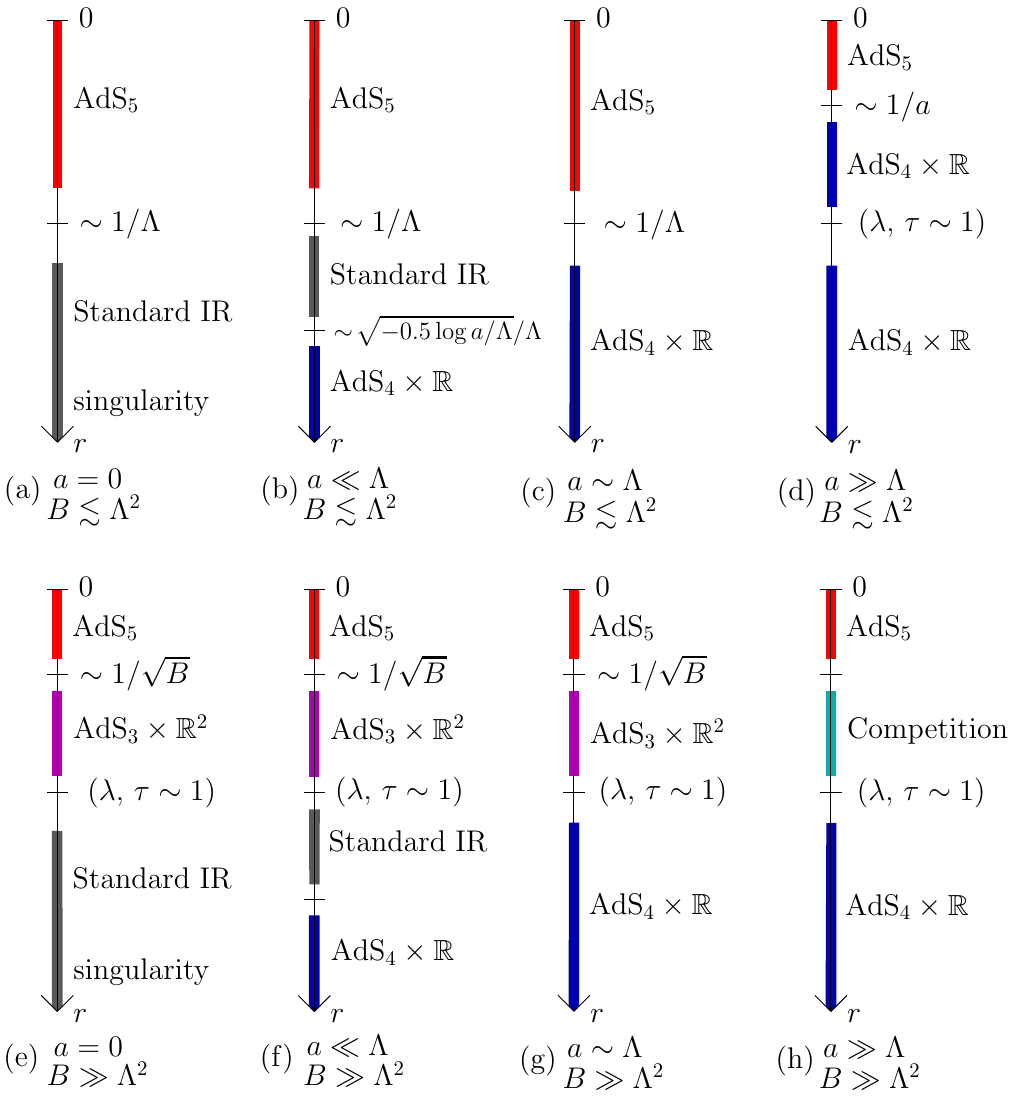}
\caption{The holographic RG flow in the model. We sketch the geometry for different values of $a$ and $B$,  as indicated in the plot, as a function of the holographic coordinate $r$ from the UV ($r = 0$) to the IR ($r = \infty$). The red, gray, blue, and magenta sections show the AdS$_5$, ``standard IR'', AdS$_4\times \mathbb{R}$, and AdS$_3\times\mathbb{R}^2$, given in~\eqref{UVexpsapp}, in~\eqref{bgstandardIR}, in~\eqref{IRasympt}, and (approximately) in~\eqref{ads3geom}, respectively. We observe that non-vanishing $a$ removes the typical singularity at the origin of the geometry ($r= \infty$) ~\cite{Gursoy:2018ydr}. Transition between different regions are marked by the value of $r$ or the scalars $\lambda$ and $\tau$. The two (approximate) AdS$_4\times \mathbb{R}$ regimes in (d) have different radii. Effects of $a$ and $B$ compete in the intermediate regime in (h). \label{fig:RG_flow}}
\end{figure}

\subsection{The overall structure of the RG flow}

The results from the above analysis can then be combined to draw the holographic RG flows for the zero temperature flows for various values of $a$ and $B$. The sketch in Fig.~\ref{fig:RG_flow}, where $B$ is the absolute value of the magnetic field, covers all possible scenarios. Notice that all of these flows are divided into separate UV sections (where the scalars $\lambda$ and $\tau$ are both $\ll 1$) and IR sections (where the scalars are both $\gg 1$). For the first three flows, (a), (b), and (c), the transition from the UV to the IR takes place at $r \sim 1/\Lambda$.

The flows on the top row are for small values of the $B$ field. In this case the magnetic field does not affect the qualitative features of the flow: the terms involving $B$ are suppressed both in the UV (where they are $\sim r^4 B^2$) and in the IR (where the growth of the tachyon decouples the flavors, including the effect of the magnetic field). Therefore the flow is qualitatively similar to the case of  zero magnetic field, studied in~\cite{Gursoy:2018ydr}. Notice the flow (d) at large $a/\Lambda \gg 1$ where the geometry has separate AdS$_4$ sections at intermediate energies and in the IR. The intermediate section is analogous to the intermediate $B$ geometry discussed in Sec.~\ref{sec:intermB}, and we have verified its existence also numerically. We note that the AdS$_4$ geometries here are always approximate: even the asymptotic IR geometry given in~\eqref{IRasympt} has multiplicative logarithmic corrections.

For large values of the magnetic field, $B/\Lambda^2 \gg 1$, an intermediate AdS$_3$ scaling discussed in Sec.~\ref{sec:intermB} emerges as sketched for the flows (e), (f), and (g) on the bottom row. We note that this (as well as the intermediate AdS$_4$ behavior of the flow (d)) happens in  the UV regime in the sense that the values of the scalars are small, $\lambda \ll 1$ and $\tau \ll 1$. The IR behavior of the geometry is independent of $B$ even at large values of the field due to the rapid decoupling of the flavors as $\tau$ grows towards the IR. We stress that the geometries are only approximately AdS$_3$: $\lambda$ is not exactly constant (which would be required for the geometry to be exactly AdS$_3$) but flows slowly. These AdS$_3$ regimes can nevertheless be clearly identified also in the numerical solutions at large $B$\@.

Finally, in Fig.~\ref{fig:RG_flow} (h) we present the flow when both $a$ and $B$ are large. In this case the AdS$_3$ and AdS$_4$ geometries ``compete'' in the intermediate section of the geometry. We have not tried to carry out a detailed analysis of this competition, in part because we are mostly interested in the behavior of the theory up to moderate values of $a$ and $B$. It also turns out that constructing the RG flow numerically when both $a$ and $B$ are large is very demanding. This does not, however, imply little competition between $a$ and $B$ in our analysis: on the contrary, as we shall see below, there are interesting interference effects in most of the observables we study. The competing results just arise from the regime where $\lambda \sim 1$ and $\tau \sim 1$. In this transition region the flow of the scalars is fast and they couple nontrivially to the metric, so the metric cannot be approximated by a simple geometry.

Notice also that the RG flows of Fig.~\ref{fig:RG_flow} only cover the flows which are chirally broken. In particular we implicitly assumed that the ratio $x_f=N_f/N_c$ is not small. All zero temperature backgrounds studied in the rest of the article (where we use $x_f=1$) fall into the classification of Fig.~\ref{fig:RG_flow}. However, in~\cite{Gursoy:2018ydr} we also found chirally symmetric vacua for $x_f=1/3$, for which the RG flow ends at a fixed point with exact AdS$_4\times \mathbb{R}$ geometry in the IR. This RG flow is roughly analogous to the UV part of the flow in Fig.~\ref{fig:RG_flow} (d). Detailed analysis of chirally symmetric flows is left for future work.\footnote{The chirally symmetric fixed point observed in \cite{Gursoy:2018ydr} entails appearance of a (quantum) critical behavior for a particular value of ``doping'', which may be interesting for condensed matter applications.}

\section{Thermodynamics}
\label{sec::thermo}

The thermodynamics of the model has been studied extensively in previous works, as a function of $T$ and the chemical potential $\mu$ for the baryon number~\cite{Alho:2012mh,Alho:2013hsa}, as a function of~$T$ and $B$ (and $\mu$)~\cite{Gursoy:2016ofp,Gursoy:2017wzz}, and as a function of $T$ and $a$~\cite{Giataganas:2017koz,Gursoy:2018ydr}.\footnote{The phase diagram in the presence of anisotropy resembles that of a charged black hole in the canonical ensemble, with a swallow tail in the free energy and an associated Van der Waals like transition \cite{Chamblin:1999tk,Caceres:2015vsa,Pedraza:2018eey}.} For vanishing $a$ a typical phase diagram contains, at low temperatures a confined thermal gas phase which is dual to a singular geometry without a horizon, and at high temperatures a deconfined phase which is dual to a black hole geometry. Chiral symmetry breaking is realized through condensation of the tachyon field in the bulk.
Typically the transition from the confined phase to the deconfined phase coincides with the restoration of chiral symmetry, but depending on the values of $B$, $x_f$, and the precise choice of potentials, deconfinement and chiral transition may also be separate \cite{Gursoy:2016ofp,Gursoy:2017wzz}\@.

For finite values of the anisotropy parameter $a$, however, the low temperature thermal gas solution is replaced by a small black hole~\cite{Giataganas:2017koz,Gursoy:2018ydr}. This structure reflects a drastic change in the IR structure of the theory: the zero temperature asymptotic IR geometry becomes AdS$_4\times \mathbb{R}$ at finite $a$. This remains true also in the presence of a magnetic field, see Sec.~\ref{app:asymptotics}. For the values of anisotropy parameter we study here (excluding the case $a=0$ which we include for reference), the geometry has a black hole at all nonzero temperatures, and the only finite temperature transition is a (second order) chiral transition realized by disappearance of the tachyon condensate in the bulk.

\subsection{Thermodynamic potentials}

We start by reviewing the relevant thermodynamic potentials and their relations (see also~\cite{Mateos:2011tv,Ammon:2012qs}). For details see Appendix~\ref{app:thermodynamics}.

First, we define magnetization $M_B$ and the analogous  quantity for anisotropy $M_a$ such that the first law of thermodynamics reads
\be \label{firstlaw}
 dF = - s dT - M_B dB - M_a da \,.
\ee
One finds the following expressions for these quantities in terms of bulk  fields~\cite{Gursoy:2016ofp,Gursoy:2018ydr}
\begin{align}
 M_B & = -M^3N_c N_f B \int_{\epsilon}^{r_h} dr  \frac{ G(r) e^{A(r)+W(r)} w(\lambda (r))^2 V_f(\lambda (r),\tau (r))}{Q(r)} \,,\label{MBdef} \\
 \label{Madef}
 M_a &= -M^3N_c^2a \int_{\epsilon}^{r_h}dr\, e^{3 A(r)-W(r)} Z(\lambda (r)) \,, &
\end{align}
where
\be
G = \sqrt{1 + e^{-2A}f\kappa(\lambda)\dot\tau^2}\,, \qquad Q = \sqrt{1 + w^2(\lambda)B^2e^{-4A - 2U}}\,.
\ee
The free energy $F$, as well as $M_B$, and $M_a$ are also subject to UV divergences and need to be renormalized~\cite{Papadimitriou:2011qb}. For our purposes it is enough to do this by subtracting a reference background -- see Appendix~\ref{app:thermodynamics} as well as~\cite{Gursoy:2016ofp,Gursoy:2018ydr} for details.

Recall that the anisotropy is either in the direction of the magnetic field ($a_\parallel$) or perpendicular to it ($a_\perp$). Let us first discuss the thermodynamics in the former case, $a=a_\parallel$. Then the (non-renormalized) free energy density may be expressed in terms of boundary and horizon data as (see Appendix~\ref{app:thermodynamics}), see appendix \ref{app:thermodynamics}.
\be
F=M^3N_c^2e^{3 A(\epsilon)+W(\epsilon)} \left(6 f(\epsilon) A'(\epsilon)+ 2f(\epsilon) W'(\epsilon)\right) -s T
\ee
where $\epsilon$ is a UV cutoff. Notice that the first term on the right hand side in this equation equals the energy density $\mathcal{E} = T_{00} = F +sT$.
Denoting the pressures, $p_i = T_{ii}$, parallel and perpendicular to the magnetic field as $p_\parallel =p_3$ and $p_\perp = p_1=p_2$, we find that
\be
 {\mathcal{E}} + p_\perp = sT
 -B {M}_B \,,\qquad {\mathcal{E}} + p_\parallel = sT
 -a {M}_a
\ee
so that the relation between the free energy and pressure is
\be
 {F} = -  p_\perp -B {M}_B =-p_\parallel-a {M}_a \,.
\ee
Moreover, we note that the pressure anisotropy can be expressed in terms of the (normalized) magnetization and $M_a$:
\be
p_\parallel- p_\perp = B {M}_B - a {M}_a \,.
\ee

In the latter case, $a=a_\perp$, the free energy density can be expressed in terms of boundary data only:
\be
 F = M^3N_c^2e^{3 A(\epsilon)+U(\epsilon)+W(\epsilon)} \left(6 f(\epsilon) A'(\epsilon)+ 2f(\epsilon) U'(\epsilon)+f'(\epsilon)\right) \,.
\ee
As the magnetic field and the anisotropic parameter act in different directions, all components of the pressure are independent. We find that
\be
  {\mathcal{E}} + p_1 = sT
  -B {M}_B \,,\qquad {\mathcal{E}} + p_2 = sT
  -B {M}_B-a {M}_a\,,\qquad {\mathcal{E}} + p_3 = sT
\ee
so that the relation between the free energy density and the pressure components reads
\be
 {F} = -  p_1 -B {M}_B =-p_2-B {M}_B-a {M}_a = -p_3 \,.
\ee
The pressure differences can again be expressed in terms of $M_B$ and $M_a$:
\be
  p_3-p_1 = B {M}_B \,,\qquad  p_2-p_1 = -a {M}_a \,.
\ee

\subsection{Phase structure and the chiral transition}
\label{sec::ps}

We now study the phase diagram when the effects of the magnetic field $B$ and the anisotropy parameter $a$ compete. We fix $x_f=1$, which corresponds to the large N analog of real QCD with $N_c=3$ colors and $N_f=3$ light flavors. Small values of the parameter $c$ give qualitative agreement with lattice QCD data in the presence of $B$ \cite{Bali:2011qj,Bali:2012zg,DElia:2012ems} which exhibits inverse magnetic catalysis as we discussed in the previous section. Specifically, we will use either of $c=0.25$ and $c=0.4$. Then (except for very small values of $a$), as mentioned above, the only nontrivial feature of the phase diagram is the chiral transition. This was shown at zero $B$ and at $c=0.4$ in~\cite{Gursoy:2018ydr}, and continues to hold at finite $B$ as well as for $c=0.25$.
We also note that in a related model~\cite{Iatrakis:2010zf,Iatrakis:2010jb}, where chiral symmetry breaking arises as in V-QCD through a tachyonic brane action but backreaction is not considered, only magnetic catalysis was found~\cite{Ballon-Bayona:2020xtf}. This result agrees with our earlier work, where backreaction was seen to be essential for inverse magnetic catalysis~\cite{Gursoy:2016ofp}. 
In the next section we explore other observables that help characterize the $c$-dependence and, possibly, help single out a preferable value once first principles (lattice-QCD) calculations of these observables become available.

We recall that the behavior of the chiral transition temperature $T_\chi$ as a function of $B$ can be used as a proxy for (inverse) magnetic catalysis\footnote{which we abbreviate by (I)MC below.}, where decrease of $T_\chi$ with increasing $B$ signals inverse magnetic catalysis \cite{Bali:2012zg}. As noted in \cite{Gursoy:2016ofp,Ballon-Bayona:2017dvv,Gursoy:2017wzz}, using the first law of thermodynamics one can rewrite the derivative $dT_\chi/dB$ in terms of differences in entropy and magnetization between the chirally symmetric and asymmetric phases. This quantity can in turn be used as an order parameter to diagnose (I)MC. One can derive an analogous result in the case of anisotropy, where inverse anisotropic catalysis can be detected by looking at the differences in entropy and `anisotropization' between the two phases\footnote{In principle, one could also derive a two-dimensional analog of these arguments, where one can see the behavior of the chiral transition temperature as a function of both $a$ and $B$. However, since we only look at slices through the phase diagram at fixed $a$, we will not explicitly perform this computation. This corresponds to specifying a particular ensemble where we fix $a$ in the dual field theory.} \cite{Gursoy:2018ydr}.

\begin{figure}[t!]
\centering
\includegraphics[width=0.5\textwidth]{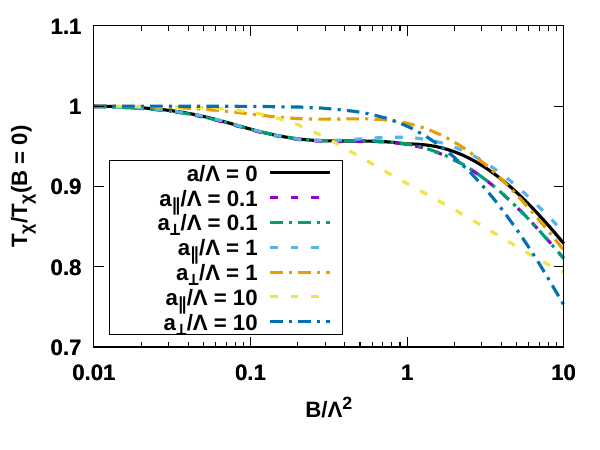}\includegraphics[width=0.5\textwidth]{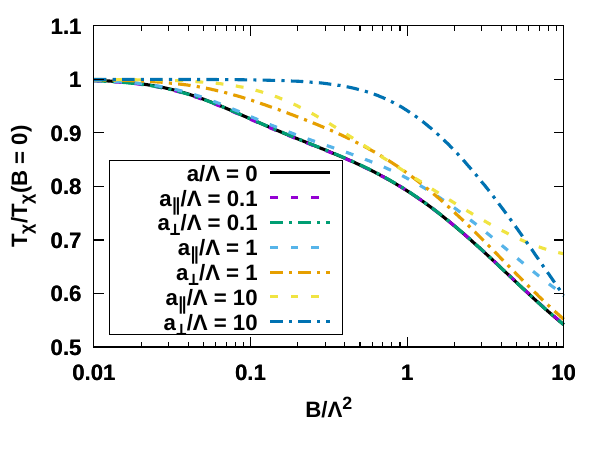}
\caption{Chiral transition temperature for $c=0.4$ (left panel) and $c=0.25$ (right panel) as a function of the magnetic field for various values of the anisotropic parameter and for different orientation choices as indicated in the legend. \label{fig:TransTemp}}
\end{figure}
The results for the chiral transition temperature are shown in Fig.~\ref{fig:TransTemp}. Each curve is normalized to the transition temperature at $B=0$. We observe that the transition temperature decreases (inverse catalysis) with increasing $B$ for almost all the curves. Inverse catalysis is slightly weaker for $c=0.4$ (left plot) than for $c=0.25$ (right plot), in agreement with the results of~\cite{Gursoy:2016ofp}.  There is a small interval at intermediate $B$ and small $a$ where the transition temperatures increase (weakly) with $B$ for $c=0.4$, thus alternating between IMC and ordinary MC. This is a feature of our model for this particular value of $c$ and it is absent for $c=0.25$. However, we still need to benchmark our model against lattice data to determine which value would be favored while comparing other observables of this model with QCD (see Section \ref{sec:otherObs} below).

For both values of $c$, the characteristic scale of $B$ for which the transition temperature starts to deviate from its $B=0$ value grows with $a$. We also note that at small $B$, curves with nonzero $a_\parallel$ are lower than the curves with nonzero $a_\perp$ (for the same numerical value of the parameter)\@. These observations are in agreement with the idea (anisotropic catalysis) that the inverse magnetic catalysis may be due to the anisotropy created by the magnetic field~\cite{Giataganas:2017koz,Gursoy:2018ydr}: the effect is stronger when the anisotropy created by the two different sources is in the same direction, and for the effect to be visible, $B$ needs to be comparable to $a$. Moreover, we observe that the transition temperatures for nonvanishing $a_\perp$ and  $a_\parallel$ with the same magnitude always cross as $B$ varies. The value of $B$ where the crossing takes place is roughly $B \sim a$, except at $a_\perp/\Lambda=0.1=a_\parallel/\Lambda$, for which the crossing (not visible in the plots) takes place at $B/\Lambda^2 \gtrsim 1$.
Finally, we note that in the $c=0.4$ case turning on anisotropy (both when parallel and perpendicular to $B$) increases (decreases) the transition temperature for small (large) $B$. This alternating behavior is absent in the case $c=0.25$ \footnote{Rather the turning point seems to be pushed to much larger values of $B$ outside the range of these plots.}.

\section{Other observables\label{sec:otherObs}}

\subsection{Quark-antiquark potential}
\label{sec::qqpot}

In this section we will study in detail an important observable of physical interest, namely the quark-antiquark potential.\footnote{Heavy quark observables were first studied in simpler anisotropic models in \cite{Giataganas:2012zy,Chernicoff:2012iq,Chernicoff:2012gu,Giataganas:2013lga}.} We emphasize  that some lattice results are available for this observable in the presence of a magnetic field \cite{Bonati:2014ksa}, providing a good opportunity to benchmark our model and study in more detail the dependence on the parameter $c$.

Akin to quantum chromodynamics, the ihQCD model is engineered to exhibit linear confinement in the low-temperature phase: the quark-antiquark potential $V(L)$ grows linearly with the separation of the probe quark-antiquark pair for large enough $L$. Such a potential acts as a barrier that prevents colored particles from escaping their bound states and become free. On the other hand, as we discussed above, anisotropic deformation in the ihQCD model substantially modifies the deep IR of the theory \cite{Gursoy:2018ydr}: the geometry becomes AdS$_4\times \mathbb{R}$ up to small corrections, indicating an emergent approximate conformal symmetry in one dimension lower. As a result, linear confinement will only persist up to a certain length scale. This is
evident from the behavior of the quark-antiquark potential, which shows signs of instability at large enough separations \cite{Gursoy:2018ydr}. Related to this, it was observed in the same reference that mesons and glueballs indeed fail to be absolutely stable but instead develop narrow widths, indicating the possibility of decaying to the AdS$_4$ vacuum.

Even though the presence of a magnetic field induces an intermediate AdS$_3 \times \mathbb{R}^2$ region (see section \ref{app:asymptotics}), the effect of $B$ on the quark-antiquark potential is qualitatively different from that of $a$\@.
The reason for this is that in our setup $B$ enters through the flavor sector, in contrast to simpler models, e.g.~\cite{DHoker:2012rlj,Rougemont:2014efa}, 
where $B$ is introduced by means of a Maxwell term. Since the flavor sector decouples in the deep IR, the effect of $B$ on the geometry is negligible, and in absence of $a$ the IR geometry is the same as when $B = 0$\@.
As a consequence of this, linear confinement is not lost in the presence of $B$\@.
The purpose of this section is to investigate the interplay between the two sources of anisotropy, $a$ and $B$, in the behavior of the quark-antiquark potential\footnote{We refer to \cite{Arefeva:2018hyo,Arefeva:2018cli} for an analysis of the potentials in slightly simpler anisotropic models.}\@.

In the context of holography, the quark-antiquark potential can be computed as a sum over saddle points, with several terms contributing \cite{Bak:2007fk}.
In the approximation that we are working in, i.e., the $\alpha' \rightarrow 0$ (infinite coupling) limit, only the term with the smallest action survives.\footnote{We neglect the graviton exchange contribution, among others, which maintains the smoothness of the Polyakov loop two-point function as a function of $L$ \cite{Bak:2007fk}. However, this contribution does not qualitatively change our results.} This term can be computed by evaluating the Nambu-Goto action on-shell, for a static string in the 5D bulk spacetime (in the string frame) with its endpoints reaching the AdS boundary \cite{Maldacena:1998im,Rey:1998ik}. We focus on string configurations with boundary conditions defining a rectangular loop with $t\in[-\frac{T}{2},\frac{T}{2}]$, $x_i\in[-\frac{L}{2},\frac{L}{2}]$ and $x_j=0$ for $j\neq i$. In the limit $T\to\infty$ the quark-antiquark potential can be extracted from
\be\label{expwl}
\int \mathcal{D}\Sigma\, e^{-S_{\text{NG}}(\Sigma)}=e^{-T V_i(L)}\,,\qquad S_{\text{NG}}=T_f\int_\Sigma \sqrt{-\det g_{ab}}\,,
\ee
where $T_f\equiv(2\pi\alpha')^{-1}$ is the string tension. Since our system is completely anisotropic, we have defined $V_i(L)$ ($i=1,2,3$) as the binding energy, or quark-antiquark potential, of the pair when they are separated along the $x_i$-direction. In principle, one could also look at the potential in arbitrary directions, but due to the lack of symmetry, the calculation would be a bit more demanding. We will therefore refrain from showing these results here. We also have two physically distinct configurations, namely when the anisotropic deformation is parallel or perpendicular to the magnetic field; we refer to these two cases as $V_i^\parallel$ and $V_i^\perp$, respectively.

Now, instead of computing $V_i$ directly as a function of $L$, in practice it is easier to obtain both $V_i$ and $L$ as functions of the worldsheet turning point $r_F$, as follows \cite{Kinar:1998vq,Gursoy:2007er}:
\begin{align} \label{Videff}
\frac{V_i(r_F)}{T_f} &= e^{2A_S(r_F)+X_i(r_F)}L(r_F) + 2\int_0^{r_F}\frac{\mathrm{d}r}{e^{X_i(r)}}\sqrt{e^{4A_S(r)+2X_i(r)} - e^{4A_S(r_F)+2X_i(r_F)}}&\nonumber\\
& \phantom{=} - 2 \int_0^{\infty}\mathrm{d}r\, e^{2 A_S(r)}\ ,&\\
L(r_F) &= 2\int_0^{r_F}\frac{\mathrm{d}r}{e^{X_i(r)}}\frac{1}{\sqrt{e^{4A_S(r)+2X_i(r)-4A_S(r_F)-2X_i(r_F)} - 1}}\ ,&\label{eq:VparL}
\end{align}
where $A_S = A + \frac{2}{3}\log\lambda$ is the string frame scale factor. We have also defined collectively $X_i(r)=\{0,U(r),W(r)\}$ for ease of notation. In general, the on-shell Nambu-Goto action is divergent, because the worldsheet reaches the boundary of AdS, where the volume element blows up. This can be easily understood from the field theory perspective, since the mass of the probe quark-antiquark pair is infinite. In order to regulate the result, we have
added the last term in~\eqref{Videff} which equals twice the action of a straight string hanging from the boundary to the IR\@. This term effectively subtracts the rest energy of two isolated quarks.

\begin{figure}[t!]
\centering
\includegraphics[width=0.50\textwidth]{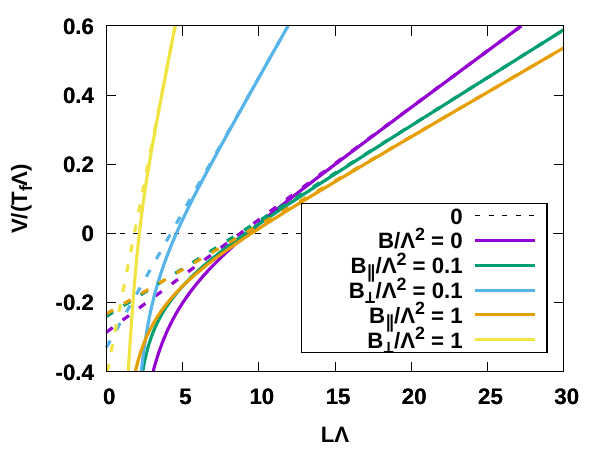}\includegraphics[width=0.50\textwidth]{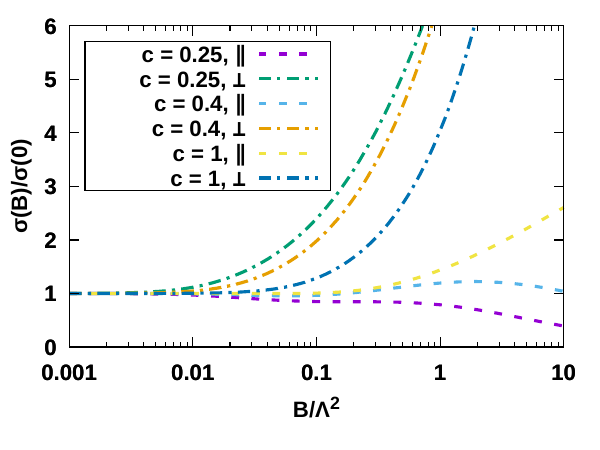}
\caption{Quark-antiquark potentials and string tensions for $a=0$. The plots present two cases, corresponding to the $q\bar{q}$ pair separated along the direction of the magnetic field ($B_\parallel$) or one of the directions orthogonal to it ($B_\perp$), all in units of $\Lambda\sim1$ GeV\@.
The potentials themselves (left panel) are shown together with the asymptotic behavior of the potentials for large separation (shown dashed)\@. The string tensions (right panel) represent the slope of the potentials in the confining regime. For these plots we varied the parameter $c$ introduced in (\ref{def:c}), which controls the region of intermediate energy scales. The $c=0.25$ case displays both IMC and agrees qualitatively with the results of \cite{Bonati:2014ksa}, in the range of magnetic fields considered there $eB\sim0-1.2$ GeV.\label{fig:qqpotB}}
\end{figure}

Following \cite{Kinar:1998vq,Gursoy:2007er}, it is easy to show that at large $L$, (a branch of) the quark-antiquark potential $V_i$ grows linearly with $L$ if $A_S + X_i/2$ has a minimum. However, while this indeed gives some indication of confinement, it does not capture the full picture. The issue is that, even for very small anisotropy, the deep IR of the theory develops an approximate AdS$_4\times \mathbb{R}$ geometry, which allows bound states to decay. One of the effects of that AdS$_4$ geometry is that one of the directions pinches off in the IR and this allows for strings to end there. This enables unbound states to exist, which means that there is always a $V_i=0$ branch that corresponds to two disconnected strings. For large enough separations, this branch will \emph{always} dominate over the linear branch, which indicates that linear confinement will only be present up to a certain length scale. The situation is quite dissimilar for the case of magnetic field only (in absence of axion anisotropy), since in that case the IR develops only an \emph{intermediate} AdS$_3\times \mathbb{R}^2$, while crucially the deep IR is unmodified. For completeness, we will show first this latter case ($a=0$, $B\neq0$), since it has not been studied so far for the ihQCD model with backreacted flavors. Results of the full computation in this case are shown in fig.~\ref{fig:qqpotB}.

In fig.~\ref{fig:qqpotB} (left panel) we have plotted the quark-antiquark potential for two cases, labeled as $B_\parallel$ and $B_\perp$. The first one corresponds to separation of the quark-antiquark pair in the same direction as the magnetic field, i.e., $V_3$, while the second one corresponds to separation orthogonal to the magnetic field, i.e., $V_{1,2}$, both with $a=0$ and $c=0.25$.
It is interesting to compare the string tensions (slopes) of this confining regime with lattice QCD calculations \cite{Bonati:2014ksa}, at least at the qualitative level. In fig.~\ref{fig:qqpotB} (right panel) we plot such tensions (normalized by the $B=0$ value) for choices of the parameter $c$. Strikingly, only the $c=0.25$ case agrees qualitatively with the results of \cite{Bonati:2014ksa} (monotonically increasing/decreasing functions of $B$ for the perpendicular/parallel cases, respectively), in the range of magnetic fields considered there $eB\sim0-1.2$ GeV\@. This qualitative agreement reinforces the comparison originally observed in \cite{Gursoy:2016ofp} based purely on IMC
and also agrees qualitatively with the results obtained in the models where magnetic field is introduced through a Maxwell term~\cite{Bohra:2019ebj,Bohra:2020qom}. 
On the other hand, it is worth noticing that for stronger magnetic fields, the string tensions for the parallel case behave non-monotonically.
It would be interesting to see whether the qualitative agreement discussed above continues to hold for larger magnetic fields as well. This requires extension of the study of \cite{Bonati:2014ksa} to lattices with larger $B$ which, on the other hand, may be technically quite challenging.
\begin{figure}[t!]
\centering
\includegraphics[width=0.85\textwidth]{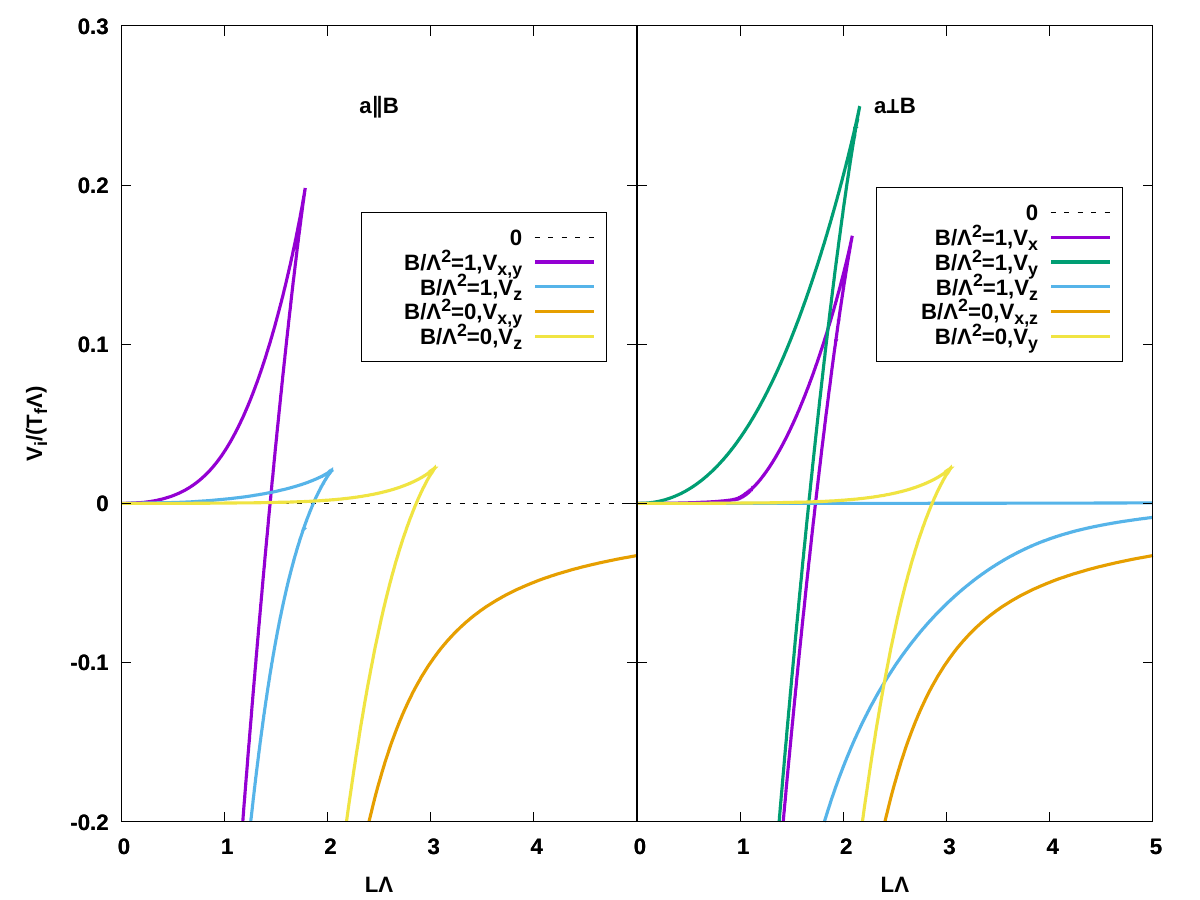}
\caption{Quark-antiquark potential in the presence of anisotropy and magnetic field, for $a=1$ and $B=1$ (in units of $\Lambda\sim1$ GeV) and $c=0.25$. The magnetic field $B$ is aligned along the $z$-direction while the anisotropic deformation is introduced along the $z$- or $y$-direction (parallel and perpendicular configurations, respectively).
In general, the combined effects of $a$ and $B$ push the transition to shorter distances, while increasing the tensions. However, two cases do not follow this pattern: $V_z$ in the case where $a$ and $B$ are perpendicular/parallel to each other. In these situations the effects of $a$ and $B$ are completely constructive/destructive in the IR and
as a result, the main effects are enhanced. This is described more in detail in the main text.
\label{fig:potBanda}}
\end{figure}
The above results should be contrasted with the behavior of the quark-antiquark potential in the presence of anisotropy (with zero magnetic field), investigated in \cite{Gursoy:2018ydr}. Surprisingly, behavior of string tensions in the parallel and perpendicular cases with non-vanishing $a$ are swapped compared to the non-vanishing $B$ case plotted in Fig.~\ref{fig:qqpotB}\@. In \cite{Gursoy:2018ydr} it was found that, in the parallel case, the slope of the potential prior to the transition is monotonic in the strength of the anisotropy $a$, while in the perpendicular case it exhibited a non-monotonic behavior, first growing and then decreasing, to ultimately disappear for strong enough anisotropy. The explanation for this change in behavior is the fact that the effects of the AdS$_4$ and AdS$_3$ geometries are exactly opposite to each other: in the anisotropic case, the dynamics are confined to the plane orthogonal to the direction of the anisotropy, while in the magnetic case, the dynamics are confined to the direction of the magnetic field.

It is then interesting to consider the case where both effects are turned on, which we investigate in fig.~\ref{fig:potBanda}. We focus on a situation where $a=1$ and $B=1$ (in units of $\Lambda\sim1$ GeV) so the two effects have similar strengths and have a chance to compete with each other. We also fix $c=0.25$ (from here on) since this is the value that seems more favorable from the comparison with the lattice \cite{Bonati:2014ksa}. We recall here that $B$ is aligned with the $z$-direction while the anisotropic deformation can be along the $z$- or $y$-directions (parallel and perpendicular configurations, respectively). In both cases we plot the quark-antiquark potential along independent directions and find a transition to the disconnected solution at large enough distances. This represents the melting of mesons at the relevant scale. In general, we observe that the combined effect of $a$ and $B$ tends to push the transition to shorter distances, while increasing the tension/slope at the moment of the transition. However, there are two cases that do not follow this pattern and are worth mentioning. The most visible counterexample is the potential $V_z$ in the case where $a$ and $B$ are perpendicular to each other. In this case, the effects of $a$ and $B$ act \emph{constructively} and strongly constrain the dynamics precisely along the $z$-direction. Instead of the normal trend, the transition in this case is pushed to larger distances, while the slope decreases significantly. The second case that does not follow the above rule, is the potential $V_z$ in the cases where $a$ and $B$ are parallel to each other. In this case the effects of $a$ and $B$ are completely opposite to each other, both trying to constrain the dynamics along different directions. The net effect in the potential is that the transition is still pushed to smaller distances but the slope/tension decreases. Altogether, we believe that these results are intuitive and physically sound; they arise due to a constructive/destructive interplay between the effects of $a$ and $B$.

\subsection{Anisotropic shear viscosity}

Besides the thermodynamic phase structure, holography is also a powerful tool to explore transport properties of QCD-like theories. We remark that first-principles methods such as lattice-QCD are not suitable to study real time dynamics at finite temperature, as computation of real-time Green's functions requires analytic continuation from the Euclidean ones, a transformation which itself requires the knowledge of the entire spectral density. In contrast, these properties can be accessed with relative ease in theories with holographic duals, often by perturbing a static solution and solving simple differential equations\footnote{Of course, one should keep in mind that holographic calculations when applied to QCD, are only meant to provide qualitative insights that are trustworthy in the IR. These are still crucial in the absence of other tools.}.

One of the most interesting observables in this context is the ratio between the shear viscosity and entropy density, which takes the universal value of $\eta/s= 1/4\pi$ for conformal plasmas at infinite coupling \cite{Policastro:2001yc,Kovtun:2003wp} and is in remarkable agreement with experimental data \cite{CasalderreySolana:2011us}. This universal value is violated in holographic theories with higher curvature gravity duals \cite{Brigante:2007nu,Brigante:2008gz,Camanho:2010ru,Cremonini:2011iq,Cremonini:2012ny} (i.e.~away from the infinite coupling limit) and even in Einstein gravity by spontaneously or explicitly breaking space-time symmetries such as translations \cite{Burikham:2016roo,Hartnoll:2016tri,Alberte:2016xja,Ling:2016ien,Baggioli:2020ljz} or rotations \cite{Erdmenger:2010xm,Rebhan:2011vd,Mamo:2012sy,Jain:2014vka,Critelli:2014kra,Jain:2015txa,Finazzo:2016mhm,Giataganas:2017koz}, 
as in our case. In all these cases the ratio exhibits interesting dependence on  temperature and other scales in the theory, which is interesting from a phenomenological point of view and reproduces better the behavior of real-world fluids.

In practice, the shear viscosity tensor can be computed via the standard Kubo formula,
\be
\eta_{ij}=-\frac{1}{\omega}\Im\, \langle T_{ij}(\omega,\vec{k}_1)T_{ij}(\omega,\vec{k}_2)\rangle\big|_{\omega\to0,\,\vec{k}_{1,2}\to0}\, ,
\ee
where the limit on the right is taken first.
For theories that break rotations, there are various independent components of the shear viscosity tensor. For example, for our specific theory,
when anisotropic deformation is perpendicular to the magnetic field, $a = a_\perp$, we find that (see Appendix~\ref{app:eta})
\begin{align}\label{etaxy}
\frac{\eta_{xy}}{s}&=\frac{1}{4\pi}\frac{g_{11}}{g_{22}}\bigg|_{r=r_h}=\frac{1}{4\pi}e^{-2U(r_h)}\,,&\\
\label{etaxz}
\frac{\eta_{xz}}{s}&=\frac{1}{4\pi}\frac{g_{33}}{g_{11}}\bigg|_{r=r_h}=\frac{1}{4\pi}e^{2W(r_h)}\,,&\\
\label{etayz}
\frac{\eta_{yz}}{s}&=\frac{1}{4\pi}\frac{g_{33}}{g_{22}}\bigg|_{r=r_h}=\frac{1}{4\pi}e^{2W(r_h)-2U(r_h)}\,.&
\end{align}
We emphasize that, although these expressions seem to naively depend only on the background geometry, they are \emph{not} general expressions that can be applied for any bulk theory: indeed the way the anisotropy is introduced does affect the formulas, in particular, giving rise to the sign difference between~\eqref{etaxy} and~\eqref{etaxz} due to our particular setting. Moreover recall that the shear viscosity tensor is symmetric, $\eta_{ij}=\eta_{ji}$. In the parallel case the result for the shear viscosities cannot be expressed in terms of background only, see Appendix~\ref{app:eta}. Since their analysis therefore needs more extensive numerical study, we leave it for future work.

\begin{figure}[t!]
\centering
\includegraphics[width=1\textwidth]{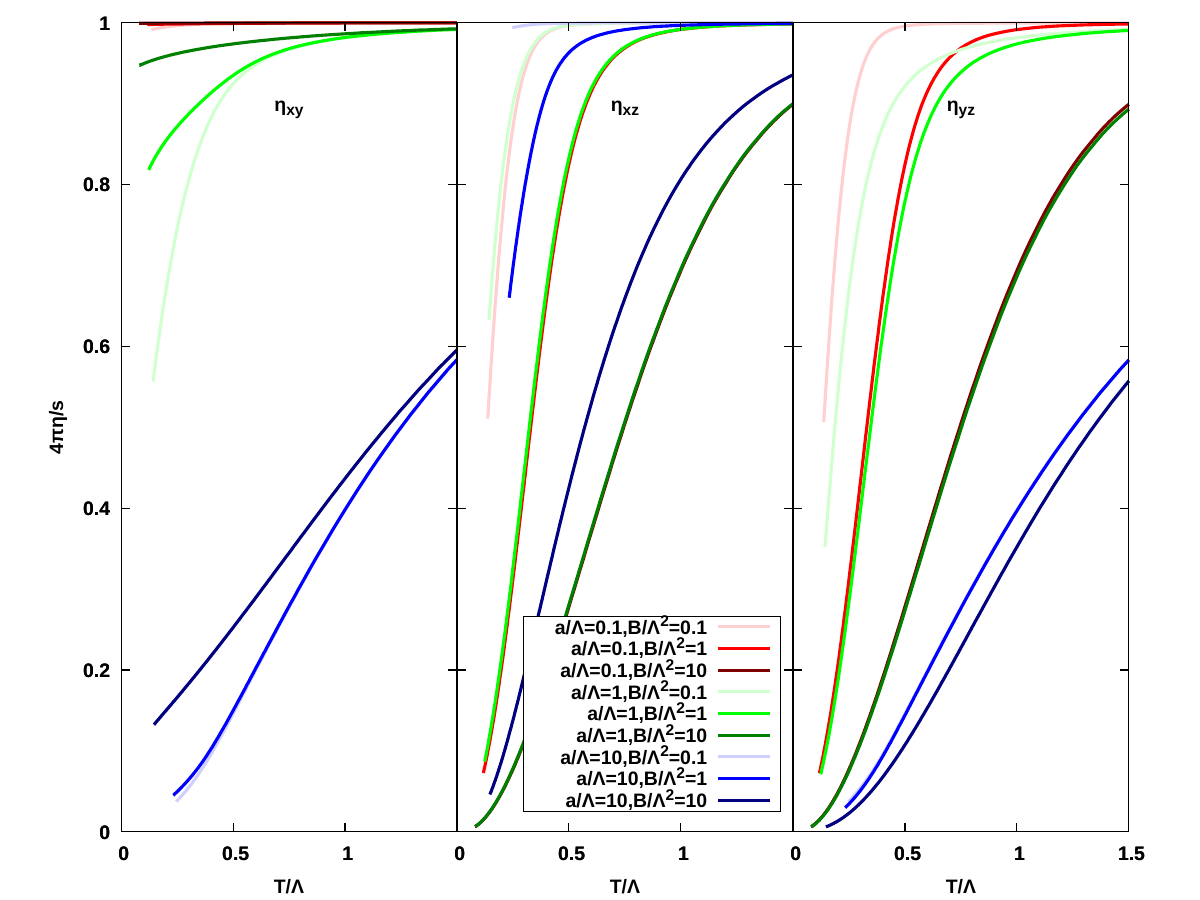}
\caption{Ratio between the components of the shear viscosity tensor and the entropy density in the chirally symmetric phase for the case where the magnetic field (aligned along the $z$-direction) is perpendicular to the anisotropic deformation (introduced along the $y$-direction)\@. All the plots are given as a function of temperature (in units of $\Lambda\sim1$ GeV) and correspond to $c=0.25$\@. Furthermore, the curves are cut off at the chiral transition temperature so they correspond to the chirally symmetric phase. In all cases, we observe a monotonic flow from $\eta_{ij}/s=1/4\pi$ in the UV to a lower, non-zero value in the IR, before the transition kicks in. We observe a clear distinction in behaviors among the different components: i) The magnetic field $B$ raises $\eta_{xy}$ and lowers $\eta_{xz}$ and $\eta_{yz}$. ii) The anisotropic parameter $a$ lowers $\eta_{xy}$ and $\eta_{yz}$ and raises $\eta_{xz}$. And iii) $B$ affects mostly $\eta_{xz}$, while $a$ affects mostly $\eta_{xy}$. The two effects are equally strong for $\eta_{yz}$ so there is no clear pattern for this component.\label{fig:shear}}
\end{figure}
In fig.~\ref{fig:shear} we plot the results of this calculation for the case when the anisotropic deformation is perpendicular to the magnetic field. As we can see, all the components of the shear viscosity tensor decrease monotonically from the UV to the IR as is the case for other anisotropic backgrounds \cite{Erdmenger:2010xm,Rebhan:2011vd,Mamo:2012sy,Jain:2014vka,Jain:2015txa,Giataganas:2017koz}. Our results show that the same trend is followed by realistic holographic models of anisotropic QCD as well. We observe that in the UV the several components attain the universal value for standard holographic CFTs, i.e., $\eta_{ij}/s=1/4\pi$, which can be explained by the fact that our solutions are asymptotically AdS\@. In the IR they attain a smaller, non-zero value. The fact that the value in the IR is non-vanishing is because these plots correspond to the chirally symmetric phase so, for every curve, we have cut the line at the chiral transition temperature. A couple of further observations are in order: first, notice that in this configuration, the effects of $a$ and $B$ should reinforce each other along the $z$- and $y$-directions but compete along the $x$-direction, as expected from the AdS$_4$ and AdS$_3$ geometries. This is indeed observed in the figures. In particular we notice that:
\begin{itemize}
  \item The magnetic field $B$ generally raises $\eta_{xy}$ and lowers $\eta_{xz}$ and $\eta_{yz}$. This can be explained by the fact that the AdS$_3$ constrains the dynamics along the $z$-direction.
  \item The anisotropy $a$ lowers $\eta_{xy}$ and $\eta_{yz}$ and raises $\eta_{xz}$. This can be explained by the fact that the AdS$_4$ constrains the dynamics along the $xz$-plane.
  \item $B$ affects mostly $\eta_{xz}$, while $a$ affects mostly $\eta_{xy}$. The two effects are equally strong for $\eta_{yz}$ and there is no clear pattern for this component.
\end{itemize}
These results are phenomenologically interesting. We have seen that, for thermodynamics and the physics of the chiral transition, $a$ and $B$ result in similar effects e.g.~IMC\@.
On the other hand, transport seems to distinguish the two, exhibiting specific behaviors as we increase/decrease one or another, with no reference to external probes. Therefore, one can argue that anisotropic shear viscosity constitutes an example of a dynamical observable that discriminates between the two origins of $SO(3)$ breaking: geometry or by charge dynamics. It would be interesting to measure the magnetic field dependence of the shear viscosity, perhaps in Bayesian analysis\footnote{The simulation still needs to be advanced by including magnetic fields.} of the heavy-ion data \cite{Novak:2013bqa,Pratt:2015zsa,Sangaline:2015isa,Bernhard:2016tnd,Bernhard:2019bmu,Devetak:2019lsk,Auvinen:2020mpc,Moreland:2018gsh,Everett:2020yty,Nijs:2020ors,Nijs:2020roc,Everett:2020xug}; or by adapting the techniques proposed in \cite{Samanta:2016pic,Samanta:2016lsh} for a condensed matter setting to the case of heavy-ion collisions.

\subsection{Entanglement entropy}

Another robust observable that can help understanding the RG flow is entanglement entropy.
Entanglement entropy is an important concept in quantum information theory that measures the amount of correlation between sub-systems in a given quantum state. It has been used extensively in quantum field theory and quantum many-body systems as a powerful tool to characterize states of matter with long range correlations, diagnose quantum phase transitions and identifying topological order. In this section, we will study the interplay between magnetic field and anisotropy in the behavior of holographic entanglement entropy, and compare the results against those for the quark-antiquark potential.

In the context of holography, the prescription to compute entanglement entropy was first proposed by Ryu and Takayanagi in their seminal paper \cite{Ryu:2006bv}. According to it, to leading order in $1/N_c$, entanglement entropy can be computed as an area,
\begin{equation}\label{rt}
S_A=\text{min}\,\frac{\mathcal{A}(\gamma_A)}{4G_N}~,
\end{equation}
generalizing the well-known Bekenstein-Hawking formula for black hole entropy. Here $S_A$ denotes the entanglement entropy of a spacelike region $A$ in the boundary theory: $S_A\equiv-\text{tr}(\rho_{A}\ln\rho_{A})$, with $\rho_{A}\equiv\text{tr}_{A^\mathsf{c}}\rho$, where $A^\mathsf{c}$ denoting the complement of $A$, being the reduced density matrix associated with $A$. The minimization in (\ref{rt}) is with respect to all areas $\mathcal{A}(\gamma_A)$ of codimension-two bulk surfaces $\gamma_A$ that are homologous to $A$ (with $\partial \gamma_A =\partial A$). Thus, similar to the quark-antiquark potential, the entanglement entropy also follows from a similar minimization procedure although in the Einstein frame, as opposed to the string frame, and of a higher dimensional surface.

The study of entanglement entropy in confining theories was first addressed in \cite{Klebanov:2007ws}. This paper proposed a generalization of the RT prescription to non-conformal field theories and found that, for gravity backgrounds which are holographically dual to confining gauge theories, the entanglement entropy generically exhibits a first order phase transition upon varying the size of the entangling surface. Typically, holographic backgrounds dual to confining gauge theories have an internal cycle that contracts smoothly and approaches zero size in the deep IR\@. This shrinking cycle leads to a ``cigar geometry'', with the IR end of space corresponding to the tip of the cigar. In these geometries, there are multiple local minima of the area $\mathcal{A}(\gamma_A)$ for a given size and shape of the region $A$. Among these, two exchange dominance as one varies the size: a connected surface that hangs at a finite radial distance from the boundary, and a disconnected one, that extends all the way to the IR (meeting at the tip of the cigar). We emphasize that the disconnected surface does in fact satisfy the homology condition, due to the existence of the contractible cycle.

We recall that the introduction of the anisotropic deformation to the ihQCD model leads to an IR geometry that takes the form of AdS$_4\times \mathbb{R}$, up to small corrections, so one of the spatial directions effectively pinches off. In terms of the entanglement entropy calculation, this has the same effect as the cigar geometry. Indeed, RT surfaces that are disconnected and reach the deep IR (Poincar\'e horizon of the AdS$_4$) effectively connect via the contracted coordinate, and therefore satisfy the homology condition. Hence, even in this theory, there are multiple branches that exchange dominance as we vary the size of the region  \cite{Gursoy:2018ydr}. We remark that the same observation holds true in the mere presence of a magnetic field, since in that situation the geometry develops an AdS$_3\times \mathbb{R}^2$ region, also reducing the dimensionality (by two in this case) in the range of relevant energy scales. See \cite{Dudal:2016joz} for a study of entanglement entropy in the presence of magnetic fields, where similar results were found. It will therefore be very interesting to study the competition of these two effects, anisotropy and magnetic field, in the behavior of the holographic entanglement entropy.

We will compute the entanglement entropy for strips $A_i$ defined as follows: $x_i\in[-\frac{L_i}{2},\frac{L_i}{2}]$ ($i=1,2,3$) and $x_j\in[-\frac{L_\perp}{2},\frac{L_\perp}{2}]$ (with $L_\perp\to\infty$) for $j\neq i$. We also have two physically distinct configurations, namely when the anisotropic deformation is parallel or perpendicular to the magnetic field; we will refer to all the possible cases as $S_{A_i,\parallel}$ and $S_{A_i,\perp}$, respectively.
We subtract the UV divergences in the same way as was done for the quark-antiquark potential, i.e., subtracting the area of two disconnected vertical surfaces.\footnote{With this regularization, we have that $S_A\to0$ as $L\to\infty$.} The final expressions for $S_{A_i}$ (both parallel and perpendicular) and $L_i$ as functions of the turning point $r_F$ are:
\begin{align}
\frac{S_{A_i}(r_F)}{4\pi L_\perp^2 M^3N_c^2} &= e^{3A(r_F)+\sum_iX_i(r_F)}L(r_F) + 2\int_0^{r_F}\frac{\mathrm{d}r}{e^{X_i(r)}}\sqrt{e^{6A(r)+2\sum_iX_i(r)} - e^{6A(r_F)+2\sum_iX_i(r_F)}}& \nonumber\\
&\phantom{=} - 2\int_0^{\infty}\frac{\mathrm{d}r}{e^{X_i(r)}} \,e^{3A(r)+\sum_{i}X_{i}(r)}\ , &
\label{eq:EE}\\
L_i(r_F) &= 2\int_0^{r_F}\frac{\mathrm{d}r}{e^{X_i(r)}}\frac{1}{\sqrt{e^{6A(r)+2\sum_iX_i(r)-6A(r_F)-2\sum_iX_i(r_F)} - 1}}\ ,& \label{eq:EE2}
\end{align}
where we have defined collectively $X_i(r)=\{0,U(r),W(r)\}$.

\begin{figure}[t!]
\centering
\includegraphics[width=0.7\textwidth]{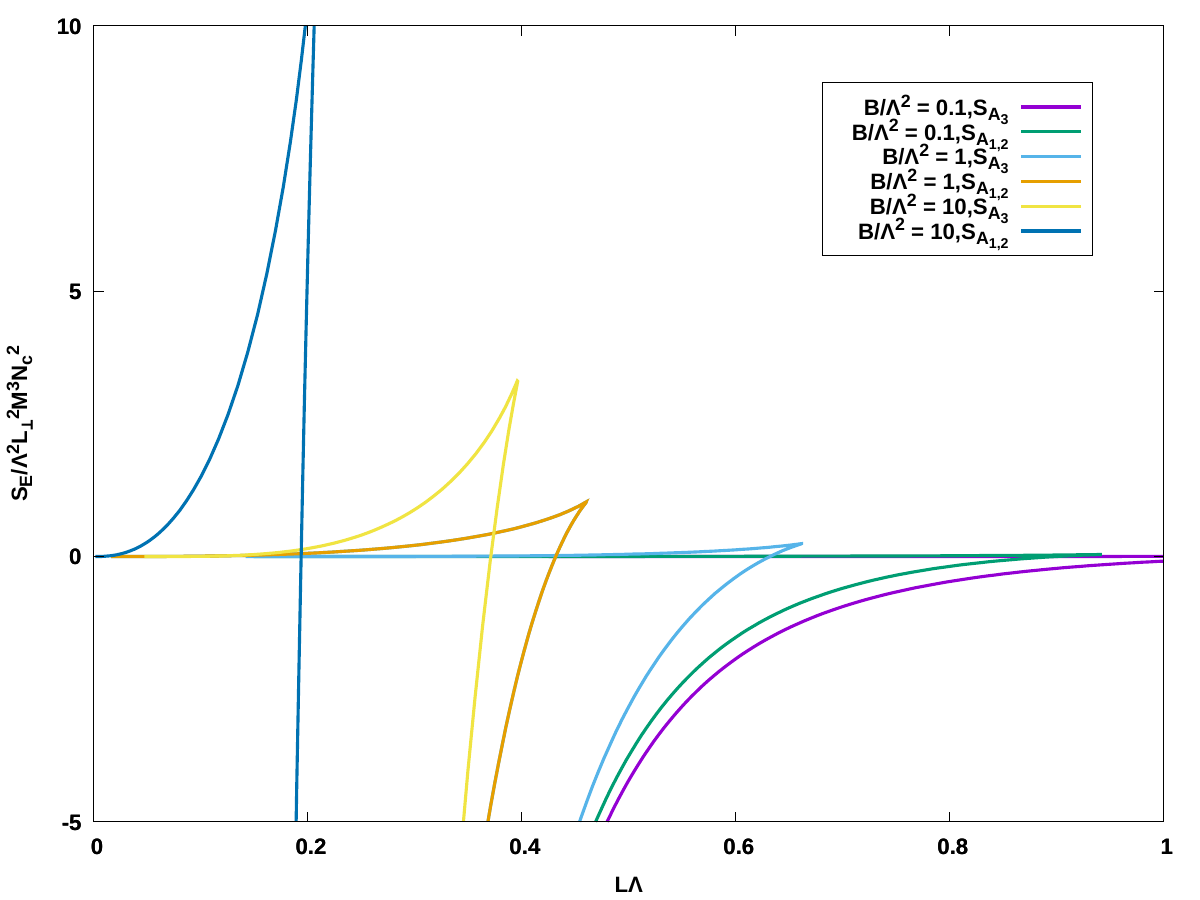}
\caption{Entanglement entropy as a function of the strip length (in units of $\Lambda\sim1$ GeV) when the anisotropic deformation is turned off ($a=0$) and $c=0.25$.
The various plots correspond to different orientations of the strip with respect to the magnetic field and different values of $B$. In all cases we observe a transition to a disconnected configuration at large distances, signaling a disentangling transition, analogous to the confinement-deconfinement transition of \cite{Klebanov:2007ws}. The effect of the magnetic field is always monotonic: as we increase $B$ the transition is pushed to smaller values of the strip length, in qualitative agreement with other holographic models with a magnetic field \cite{Dudal:2016joz}. \label{fig:entent}}
\end{figure}
\begin{figure}[t!]
\centering
\includegraphics[width=0.85\textwidth]{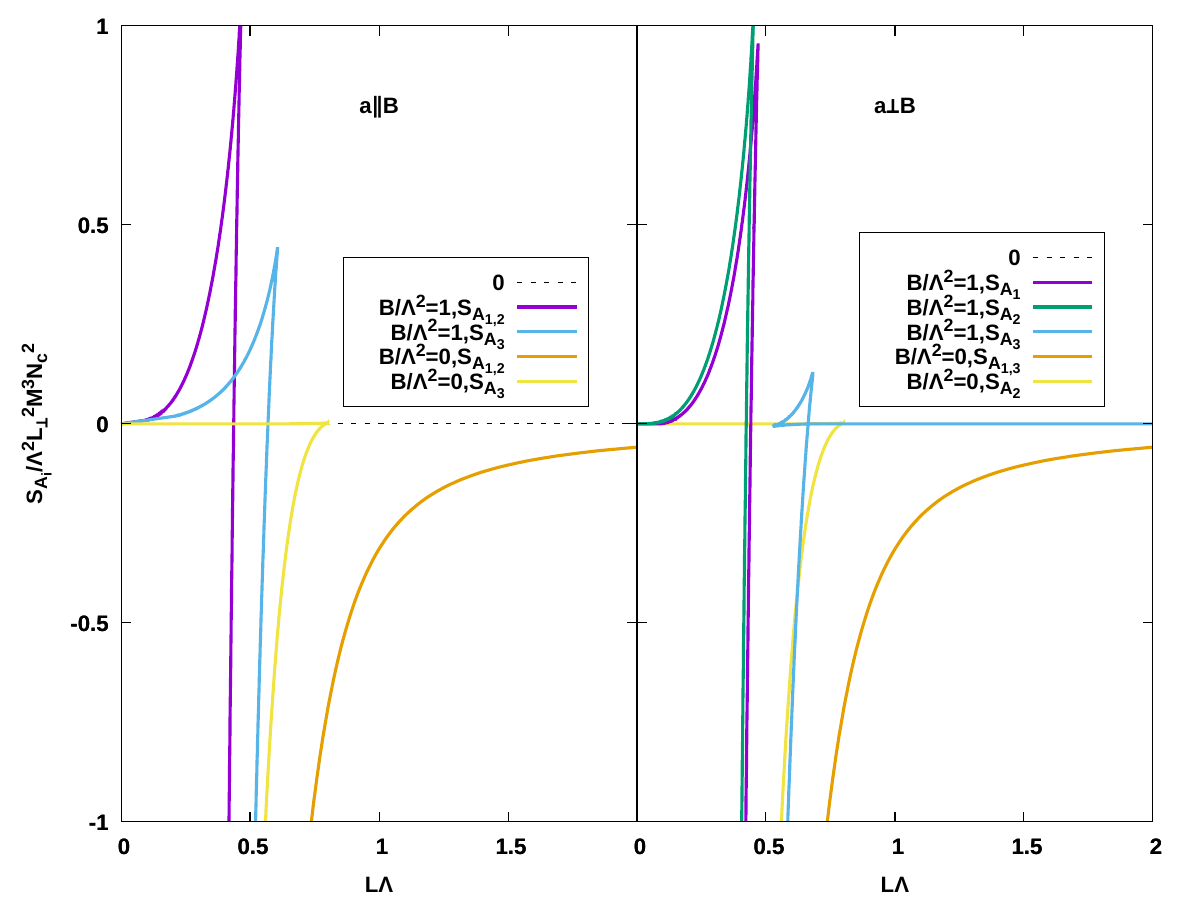}
\caption{Entanglement entropy in the presence of both anisotropy and magnetic field, for $a=1$ and $B=1$ (in units of $\Lambda\sim1$ GeV), and $c=0.25$. The magnetic field $B$ is aligned along the $z$-direction while the anisotropic deformation is introduced along the $z$- or $y$-direction (parallel and perpendicular configurations, respectively). For the parallel case, we still find the disentangling transition regardless the orientation of the strip, which is inherited from the purely magnetic case. Conversely, for the perpendicular case, the transition is replaced by swallow tail behavior and the entanglement entropy is never dominated by the disconnected solution. This behavior is inherited from the purely anisotropic case \cite{Gursoy:2018ydr}, specifically, for the case of strips orthogonal to the anisotropic deformation. Thus, the effects of $a$ and $B$ dominate depending on the physical configuration (parallel or perpendicular case).\label{fig:entBanda}}
\end{figure}
The results of these computations are shown in figures \ref{fig:entent} and \ref{fig:entBanda} below. In fig.~\ref{fig:entent} we show the result for the entanglement entropy in the presence of a magnetic field $B$ but without the anisotropic deformation, $a=0$. Generically, we observe that at very large distances or, equivalently, when the RT surface probes the AdS$_3$ region (where two of the directions pinch off), there is a transition to the disconnected solution, signaling a disentangling transition. This is equivalent to the confinement-deconfinement transition as in \cite{Klebanov:2007ws}. The results smoothly interpolate between the case without magnetic field and large magnetic field, with the increase in $B$ pushing the transition to smaller distances, in qualitative agreement with the results of \cite{Dudal:2016joz} in a simpler model with a magnetic field
(see also~\cite{Dudal:2018ztm}).
This happens for both the parallel and perpendicular configurations, as opposed to the case with only anisotropic deformation and no magnetic field \cite{Gursoy:2018ydr}, where the result strongly depends on the orientation of the strip. This can be explained by the fact that increasing $B$ moves the AdS$_3$ region more towards the UV, while the effects of $a$ are constrained more or less in the same IR region. In this latter case, it was found that for a strip aligned perpendicular to the anisotropy, the transition eventually disappears for large enough $a$. Thus, it is interesting to study the interplay between $a$ and $B$ in our model. In fig.~\ref{fig:entBanda} we consider a situation where $a=1$ and $B=1$ (in units of $\Lambda\sim1$ GeV), so the two effects have similar strengths and can compete with each other. We recall here that $B$ is aligned along the $z$-direction while the anisotropic deformation can be along the $z$- or $y$-direction (parallel and perpendicular configurations, respectively). In both cases we plot the entanglement entropy along the independent directions and observe distinct behaviors depending on the configuration. For the parallel case, we still find the disentangling transition regardless of the orientation of the strip. This is opposed to the behavior we find for the perpendicular case, where the transition is replaced by swallow tail behavior and the entanglement entropy is never dominated by the disconnected solution. This behavior is inherited from the purely anisotropic case \cite{Gursoy:2018ydr}, specifically, for the case of strips orthogonal to the anisotropic deformation.

\subsection{Butterfly velocity}

Another interesting observable that has been borrowed from quantum information theory and that can help to characterize the RG flow of the theory is the so-called butterfly velocity $v_B$. This quantity appears in the study of many-body quantum chaos and measures the response of the system to local perturbations. Specifically, $v_B$ is defined through the commutator \cite{larkin}
\be
C(t,\vec{x})=- \langle [W(t,\vec{x}),V(0,0)]^2 \rangle\,,
\label{eq-C(t,x)}
\ee
which, for chaotic systems, is expected to grow as
\begin{equation}
C(t,\vec{x}) \sim \frac{1}{N^2} \exp \left[ \lambda_L\left(t-\frac{|\vec{x}|}{v_B}\right) \right]\,, \qquad |\vec{x}|>\!\!>\beta\,,\quad t>\!\!>\beta \,,
\label{eq-C(t,x)-2}
\end{equation}
for generic hermitian operators $W$ and $V$. The quantum Lyapunov exponent $\lambda_L$ diagnoses fast scrambling, and has an upper bound in general quantum systems \cite{Maldacena:2015waa},
\be
\lambda_L\leq\frac{2\pi}{\beta}\,.
\ee
Remarkably, this bound is saturated for quantum field theories with gravity duals\footnote{Owing to open-closed string duality, that the bound is also saturated in the open string sector \cite{deBoer:2017xdk,Murata:2017rbp,Banerjee:2018twd}.} \cite{Shenker:2013pqa,Shenker:2014cwa} as well as ensemble theories
such as the Sachdev-Ye-Kitaev model and its cousins \cite{Maldacena:2016hyu,Gu:2016oyy,Davison:2016ngz}. The butterfly velocity $v_B$ characterizes the rate of expansion of $V$ in space, due to a local perturbation caused by $W$. This quantity defines an emergent light cone $\Delta t = |\vec{x}|/v_B$ such that within the cone $C(t,\vec{x}) \sim \mathcal{O}(1)$, whereas outside the cone $C(t,\vec{x}) \approx 0$. Based on this observation, \cite{Roberts:2016wdl} argued that, in holographic theories, $v_B$ acts as a low-energy Lieb-Robinson velocity $v_{LR}$ which sets a bound for the rate of transfer of quantum information.

Now, for theories that are anisotropic the emergent light cone is not expected to be invariant under rotations. In fact, the butterfly velocity defines now a vector $v_B^i$ that constrains the propagation of information along the several directions. It is interesting to note that anisotropy can in certain cases enhance the effective light-cone, at least along a subset of the components of $v_B^i$ \cite{Giataganas:2017koz,Jahnke:2017iwi,Avila:2018sqf}, thus, improving the efficiency  rate for transfer of quantum information attained in isotropic theories. It will therefore be very interesting to compute this quantity in our model and study how it behaves in the IR\@. In particular, it will be enlightening to understand the dependence on the two different sources of anisotropy that we consider, the axion deformation $a$ and the magnetic field $B$. We note that this will be the first calculation of this type in a realistic model of anisotropic holographic QCD.

The formula for the components of the butterfly velocity $v_B^i$ in general anisotropic theories is derived in detail in Appendix \ref{app:vb}. Here we will only write the final result, equation (\ref{eq:finalvb}), which specialized to our metric ansatz (\ref{ansatz}) yields:
\be\label{eq:vbour}
v_B^i=\sqrt{\frac{f'(r_h)e^{-2X_i(r_h)}}{2 [3 A'(r_h)+U'(r_h)+W'(r_h)]}}\,,
\ee
where $X_i(r)=\{0,U(r),W(r)\}$.

\begin{figure}[t!]
\centering
\includegraphics[width=1\textwidth]{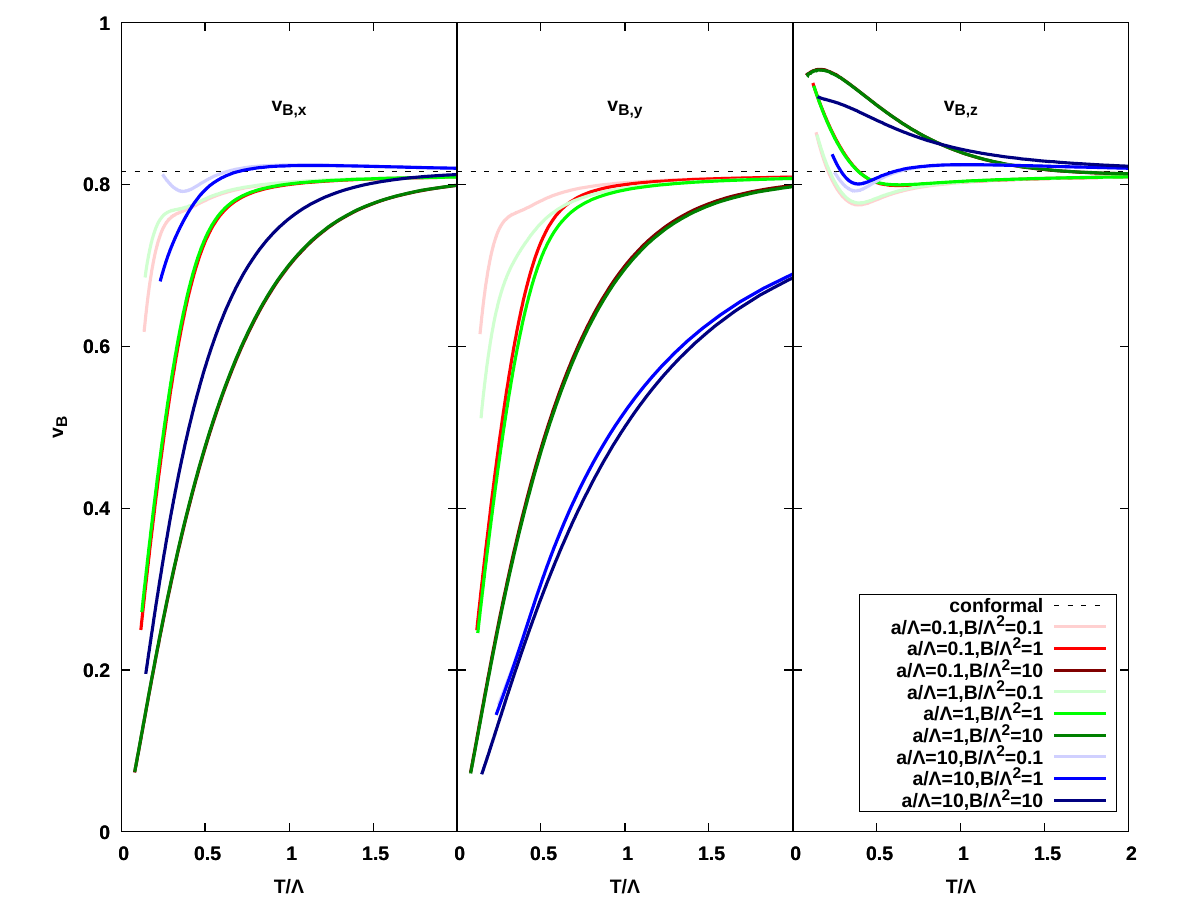}
\caption{Components of the butterfly velocity as a function of temperature (in units of $\Lambda\sim1$ GeV) for the perpendicular configuration. For all the plots we have fixed $c=0.25$. The magnetic field $B$ is aligned along the $z$-direction while the anisotropic deformation is introduced along the $y$-direction. The component $v_B^z$ is enhanced in the IR and exceeds the bound found in \cite{Giataganas:2017koz}. This is due to a constructive effect between $a$ and $B$.\label{fig:bvperp}}
\end{figure}
\begin{figure}[t!]
\centering
\includegraphics[width=0.85\textwidth]{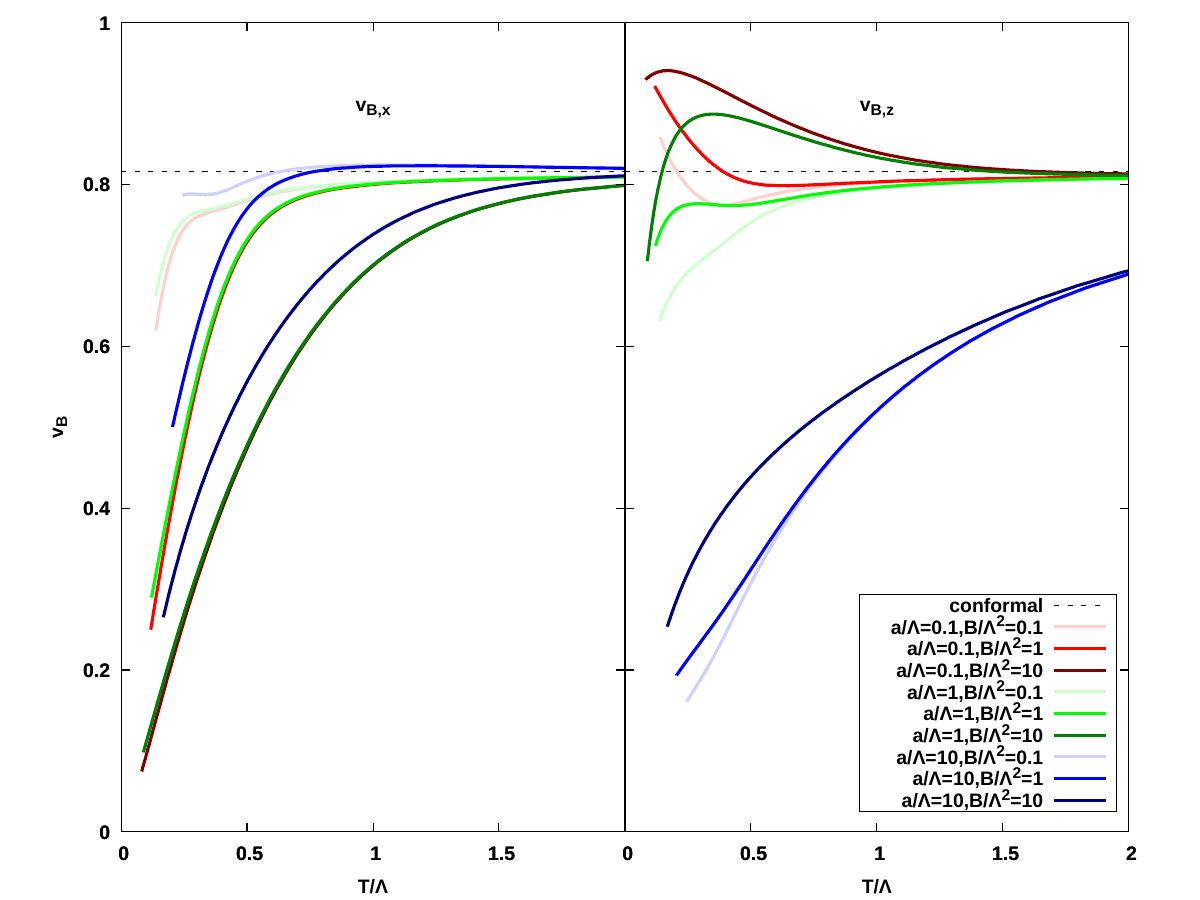}
\caption{Components of the butterfly velocity as a function of temperature (in units of $\Lambda\sim1$ GeV) for the parallel configuration. For all the plots we have fixed $c=0.25$, but in this case both the magnetic field $B$ and the anisotropic deformation are introduced along the $z$-direction. The component $v_B^z$ is also enhanced in the IR and exceeds the bound found in \cite{Giataganas:2017koz}, although it is slightly smaller than in the perpendicular case. This is due to a constructive effect between $a$ and $B$.\label{fig:bvpara}}
\end{figure}
The results of these computations are shown in figures \ref{fig:bvperp} and \ref{fig:bvpara}, for the perpendicular and parallel configurations, respectively. The various plots correspond to the independent components of $v_B^i$ in the different configurations, as a function of the temperature (in units of $\Lambda\sim1$ GeV)\@. All of them approach the conformal value in the UV, $v_B=\sqrt{2/3}\approx 0.816$, which can be explained by the fact that the geometries are all asymptotically AdS\@. However, they flow to a different value in the IR that can be either smaller or higher than the conformal value. It is interesting to note that the flow to the IR is in some cases non-monotonic, which clearly shows the competition between $a$ and $B$ at different energy scales. On general grounds, we expect that in the deep IR (specifically, before the chiral transition) the result should be mostly dominated by $a$; this is because the most relevant effect in the IR is the anisotropy, as explained in section \ref{app:asymptotics}\@. However, the non-monotonicity seems to indicate that at intermediate energies $B$ can have a dominant effect in some cases. This can be explained by the appearance of the AdS$_3$ geometry that develops because of $B$\@. It is also interesting to point out that $v_{B}^z$ (in both cases, parallel and perpendicular) can exceed the conformal value in the IR, attaining a maximum value of $v_{B}^z\approx 0.942$ in the perpendicular case. This is precisely the configuration where both $a$ and $B$ act constructively. In the parallel case the maximum is not much lower, which makes sense since $v_B^z$ attains its highest value in a regime where $B \gg a\Lambda$\@. We also note that these maxima do not appear in the deep IR: the curves have a small bump so the maximum is attained at some finite value of $T/\Lambda$ before the chiral transition.
We emphasize that this maximum is slightly higher than the upper bound found in \cite{Giataganas:2017koz}, $v_B^i\leq \sqrt{3/4}\approx 0.866$, for scaling geometries with the same kind of anisotropic deformation. This implies that the combined effect of $a$ and $B$ does indeed act constructively to raise this value. The violation of this bound is \emph{not} inconsistent with \cite{Giataganas:2017koz}, because our metric does not belong to the same universality class. It would be interesting, however, to derive new bounds in scaling geometries of the type of \cite{Giataganas:2017koz} where both the anisotropic deformation and the magnetic field are turned on \cite{Giataganas:progress}.

\section{Discussion and outlook}
\label{sec::discuss}

In this article, we studied the interplay between anisotropy and magnetic field in QCD by employing gauge/gravity duality. We considered  a setup, where we can control the effects due to the magnetic field and anisotropy separately by tuning two different sources: an anisotropic parameter $a$ and the external magnetic field $B$\@. Non-central heavy-ion collisions create quark-gluon plasma both with a high degree of anisotropic pressure gradients and large magnetic fields. Distinguishing the effects of these two sources is important both for the physics related to chiral transport that is expected to be realized in heavy ion collisions \cite{Kharzeev:2015znc,Shi:2017cpu} and for the phenomena observed in QCD in anisotropic states such as inverse magnetic catalysis \cite{Bali:2012zg}, i.e., suppression of the chiral condensate with increasing magnetic field near the deconfinement crossover temperature.

When inverse magnetic catalysis was found in lattice simulations, it came as a surprise, and its dynamical origin still remains mysterious. It was conjectured in~\cite{Giataganas:2017koz}, based on a simple model calculation, that the inverse catalysis is not due to the magnetic field directly, but rather an effect arising from the anisotropy created by the magnetic field. We studied this in a holographic setup (V-QCD) tuned to match closely with general features of QCD, and in particular the lattice data for the inverse catalysis. In~\cite{Gursoy:2018ydr} we showed that it is possible to create a similar effect, ``inverse anisotropic catalysis'' by introducing an anisotropy through a bulk axion field but in the absence of a magnetic field, therefore providing convincing evidence for the conjecture of~\cite{Giataganas:2017koz}.

In the current article, we carried out a detailed analysis of the more demanding and rich case where the system is coupled both to the anisotropic axion field and the external magnetic field. Apart from basic thermodynamics, we analyzed several observables probing the interplay of the anisotropy and the magnetic field. This provided, among other things, additional evidence for the conjecture that the inverse magnetic catalysis is an effect due to the anisotropy caused by the magnetic field: our findings for the dependence of the chiral transition temperature on the magnetic field and anisotropy parameter seems to support this conjecture.

Interestingly, gauge/gravity duality also provides a geometric interpretation for the interplay between the two effects (finite $B$ and finite $a$). As we showed in~\cite{Gursoy:2018ydr}, at finite $a$ the IR geometry is approximately AdS$_4 \times \mathbb{R}$\@. In this article we demonstrated that at large magnetic fields, the geometry contains a section which is roughly AdS$_3 \times \mathbb{R}^2$. This section is analogous to the geometry found in simpler models based on Einstein-Maxwell actions \cite{DHoker:2012rlj}. There is, however, one important difference: in the Einstein-Maxwell models the $B$ field couples directly to the geometry, whereas in V-QCD, it couples through the flavor sector, and the whole flavor sector is suppressed in the IR for the vacuum geometries due to chiral symmetry breaking. Consequently our AdS$_3 \times \mathbb{R}^2$ appears in the middle of the flow, whereas for the Einstein-Maxwell models one obtains this geometry asymptotically in the IR\@. Depending on the values of $a$ and $B$, combinations of these AdS$_3$ and AdS$_4$ geometries lead to rich possibilities for the complete holographic RG flow, as we have discussed in Sec.~\ref{app:asymptotics}.

The geometry and the RG flow are probed by the various observables which we have analyzed. Perhaps the best example of this is the quark-antiquark potential, which is found by computing the Nambu-Goto string action for a string hanging between two heavy quark sources. Notice that the directions in which the sources, $a$ and $B$, create anisotropy can be chosen independently, which leads to a further rich set of possibilities. We considered the cases where the anisotropies are in the same ($a_\parallel$) or orthogonal ($a_\perp$) directions. Notice that there is an interesting difference between $a$ and $B$ which is particularly clear in terms of the geometries. Namely, for example in the parallel case $a_\parallel$, the flat ($\mathbb{R}$) direction of the AdS$_4\times \mathbb{R}$ geometry is the $z$-direction whereas the flat directions for the AdS$_3\times \mathbb{R}^2$ geometry are $x$ and $y$. Therefore, somewhat counterintuitively, in order to obtain strong interference one needs to look at the case $a_\perp$. This is exactly what we observed: the quark-antiquark potential in the $z$-direction showed a very strong interference effect.

We also studied a closely related observable,  the entanglement entropy for strips with various alignments. Similar interference effects were observed, albeit clearly weaker. This is in agreement with the observation we made in~\cite{Gursoy:2018ydr}: The string configurations for the quark-antiquark potential extend further in the IR than the minimal surface for an analogous entanglement entropy. Therefore the quark-antiquark potential is more sensitive to the IR geometry than the entanglement entropy and a better probe for IR phenomena (see also~\cite{Jokela:2020wgs}).

Other observables we explored in this article include the shear viscosity and the butterfly velocity. This fills some gap in the literature in which these observables have not been studied in a realistic and well motivated anisotropic holographic model for QCD\@.
For the shear viscosity we obtained values which are, for some components, lower than the ``universal'' value of $s/4\pi$ even by an order of magnitude at temperatures close to the chiral transition temperature. We intuitively understand this by the fact that response of the plasma to a shear deformation is smaller in an already anisotropic state, that is response to a linear anisotropic perturbation is easily ``screened''  in an already anisotropic state. However we cannot, at this moment, provide a robust argument to support this intuition. An important observation was that while (for non-zero $a$) $a$ and $B$ affect the transition temperatures and quark-antiquark potentials in a qualitatively similar way, they have clearly distinct effects on the shear viscosity. The anisotropy affects strongly the butterfly velocities. With the $B$ field in the $z$-direction, we found that both $v_{B}^x$ and  $v_{B}^y$ typically run towards parametrically smaller values at low temperatures, well below the conformal value of $\sqrt{2/3} \approx 0.816$ for both the parallel and perpendicular configurations. For $v_{B}^z$, however, we found an interesting non-monotonic behavior, reaching a maximum of $v_{B}^z \approx 0.942$ in the perpendicular case, thus exceeding the conformal value. We note that this maximum is above the bound $v_{B} \leq \sqrt{3/4} \approx 0.866$ found in~\cite{Giataganas:2017koz}, however, this can be explained because our IR metric does not belong to the class of metrics considered in this reference.

Our results are particularly sensitive to a parameter, $c$, that enters in the flavor sector of the holographic model, see eqs.~(\ref{actf}) and  (\ref{def:c}). In particular we compared  how the phase structure and the quark-antiquark potential are influenced for two different choices $c=0.4$ and $c=0.25$ in sections \ref{sec::ps} and \ref{sec::qqpot}\@. The reason we chose these observables to investigate the $c$ dependence was motivated by the availability of lattice data \cite{Bali:2011qj,Bali:2011uf,Bonati:2014ksa} for comparison. We concluded that the choice for $c=0.25$ agrees better with lattice findings. In particular the non-monotonic dependences of the chiral transition temperature and the string tension on $B$ which arise for $c=0.4$ are absent both in the lattice studies and for our choice of $c=0.25$. Therefore, we set $c=0.25$ in our study for the rest of the observables, i.e.~the entanglement entropy and the butterfly velocity. We caution however that an exhaustive comparison of holography and lattice requires the inclusion of other observables such as the hadron spectra, Euclidean correlators, etc., and this is expected to modify the choice of parameters in our potentials.

Let us comment on other possible future directions. An extension in a somewhat orthogonal direction would be to understand better the chirally symmetric vacuum which was found for $x_f=1/3$ and for a range of values of $a$ in~\cite{Gursoy:2018ydr}. This is a fixed point solution where the dilaton runs to a constant value in the IR while the geometry is asymptotically AdS$_4 \times \mathbb{R}$ (without the logarithmic corrections of~\eqref{IRasympt}). That is, there is a quantum critical region and two zero temperature chiral phase transitions as $a$ varies. The nature of these phase transitions merits further study. It is tempting to interpret $a$ as a ``doping'' parameter and study if this configuration can be applied to condensed matter, where quantum critical regions/points are observed e.g.~in the context of high temperature superconductors.

From a general point of view, we note that despite the fact that the holographic model considered in this article is one of the most realistic holographic approaches available for QCD --- that compares well with QCD data in the various sectors \cite{Gursoy:2008cmu,Gursoy:2016ebw} --- there is room for improvement. For applications at finite magnetic field, it would be important to implement up and down type quarks in the model, so that the magnetic field would couple to the physical charge rather than baryon number.
As discussed above, there is also additional lattice data available at finite magnetic field which could be compared to the predictions of the model more precisely. Notice that after implementing the correct coupling of the magnetic field to the flavor sector, it would make sense to carry out precise quantitative fits (e.g.~to the equation of state and to quark-antiquark potentials) rather than the qualitative comparisons done so far. We also note that incorporation of anisotropy in the holographic model is done here through adding an axion field which corresponds to the topological charge operator. To complete this topological charge sector in the holographic dual, in particular to realize invariance under extended chiral transformations (that simultaneously shifts $\theta$ and an external axial gauge field) one needs to add a St\"uckelberg scalar \cite{Arean:2016hcs} which we omitted, as we are only interested in realizing the anisotropic state in this paper.

\paragraph{Note added:} While we were at the final stages of writing up this article we became aware of \cite{Arefeva:2020vae} which also considers the combined effects of magnetic field and anisotropy in a holographic model of large-$N$ QCD. Their study is based on a slightly simpler bulk theory based on an Einstein-Maxwell-Dilaton action, however, they also analyze the effects of a finite chemical potential.

\section*{Acknowledgements}

UG and GN are partially supported by the Netherlands Organisation for Scientific Research (NWO) under the VIDI grant 680-47-518, and the Delta-Institute for Theoretical Physics ($\D$-ITP), both funded by the Dutch Ministry of Education, Culture and Science (OCW). The work of MJ is supported in part by a center of excellence supported by the Israel  Science Foundation grant \#2289/18. The research of MJ is also supported by an appointment to the JRG Program at the APCTP through the Science and Technology Promotion Fund and Lottery Fund of the Korean Government.
In addition, the research of MJ is supported by the Korean Local Governments -- Gyeongsangbuk-do Province and Pohang City.
GN is supported by the U.S. Department of Energy, Office of Science, Office of Nuclear Physics under grant Contract Number DE-SC0011090.
JFP is supported by the Simons Foundation through \emph{It from Qubit: Simons Collaboration on Quantum Fields, Gravity, and Information}.

\appendix

\section{Details of the model}

\subsection{Equations of motion\label{App:eqns}}
\subsubsection{Field equations}
First, we define the following quantities:
\be
Q = \sqrt{1 + w^2(\lambda)B^2e^{-4A - 2U}}\,,
\ee
\be
G = \sqrt{1 + e^{-2A}f\kappa(\lambda)\dot\tau^2}\,.
\ee
To unify the notation between the parallel and perpendicular cases, we define $\chi = a\sin(\theta)x_2 + a\cos(\theta)x_3 = a_2x_2 + a_3x_3$, so that $a_2\equiv a\sin(\theta)$ and $a_3\equiv a\cos(\theta)$. In practice we set one of the two $a_i$'s to zero (i.e., we choose either $\theta=0$ or $\theta=\pi/2$). We further define:
\be
Y_2 = a_2^2e^{-2A - 2U}, \qquad Y_3 = a_3^2e^{-2A - 2W}, \qquad Y = Y_2 + Y_3\,.
\ee
With these notations in mind, we can now write down the Einstein's equations as follows:
\be
\ddot A + \dot A\left(3\dot A + \dot U + \dot W + \frac{\dot f}{f}\right) - \frac{e^{2A}V_g(\lambda)}{3f} + \frac{e^{2A}x_fV_f(\lambda,\tau)}{6QGf}\left(Q^2 +2Q^2G^2- G^2\right) = 0\,,
\ee
\be \label{Ueq}
\ddot U + \dot U\left(3\dot A + \dot U + \dot W + \frac{\dot f}{f}\right) + \frac{e^{2A}Y_2Z(\lambda)}{2f} = 0\,,
\ee
\be \label{Weq}
\ddot W + \dot W\left(3\dot A + \dot U + \dot W + \frac{\dot f}{f}\right) - \frac{e^{2A}Gx_fV_f(\lambda,\tau)\left(Q^2 - 1\right)}{2Qf} + \frac{e^{2A}Y_3Z(\lambda)}{2f} = 0\,,
\ee
\be
\ddot f + (3\dot A + \dot U + \dot W)\dot f - \frac{x_fV_f(\lambda,\tau)e^{2A}G\left(Q^2 - 1\right)}{Q} = 0\,,
\ee
\begin{align}
& \frac{2}{3}\frac{\dot\lambda^2}{\lambda^2} - 6\dot A^2 - 3\dot A\left(\dot U + \dot W\right) - \dot U\dot W - \frac{\dot f}{2f}\left(3\dot A + \dot U + \dot W\right) \nonumber\\
& \qquad\qquad\qquad+ \frac{e^{2A}V_g(\lambda)}{2f} - \frac{e^{2A}YZ(\lambda)}{4f} - \frac{e^{2A}x_fV_f(\lambda,\tau)Q}{2fG} = 0\,,
\end{align}
The equations for the dilaton $\lambda$, and the tachyon $\tau$, give respectively:
\begin{align}
0 & = \ddot\lambda - \frac{\dot\lambda^2}{\lambda} + \dot\lambda\left(3\dot A + \dot U + \dot W + \frac{\dot f}{f}\right) + \frac{3e^{2A}\lambda^2}{8f}\left(\partial_\lambda V_g(\lambda) - \frac{Y}{2}\partial_\lambda Z(\lambda)\right) - \frac{3e^{2A}x_fV_f(\lambda,\tau)\lambda^2}{8f}\times\nonumber\\
& \quad \left[GQ\partial_\lambda\log V_f(\lambda,\tau) + \frac{Q(G^2-1)}{2G}\partial_\lambda\log\kappa(\lambda)+\frac{G( Q^2-1)}{Q}\partial_\lambda\log w(\lambda)\right]
\end{align}
and
\begin{align} \label{taueom}
0 & = \ddot\tau - \frac{e^{2A}G^2}{f\kappa(\lambda)}\partial_\tau\log V_f(\lambda,\tau)+ G^2\dot\tau\bigg[\left(1 + \frac{(G^2 - 1)}{G^2} + \frac{2}{Q^2}\right)\dot A+\frac{\dot U}{Q^2} + \dot W+ \frac{(G^2+1)}{2G^2}\frac{\dot f}{f}\nonumber\\
& \quad + \dot\lambda\left(\partial_\lambda\log V_f(\lambda,\tau) + \frac{(G^2 + 1)}{2G^2}\partial_\lambda\log\kappa(\lambda) + \left(\frac{Q^2 - 1}{Q^2} \right)\partial_\lambda\log w(\lambda)\right)\bigg].
\end{align}
\subsubsection{Scaling symmetries}
There are four scaling symmetries, useful to solve numerically the above system:
\begin{itemize}
\item Scaling of $A$:
\be
A \mapsto A + \delta_A, \quad r \mapsto re^{-\delta_A}, \quad
B \mapsto Be^{2\delta_A}, \quad a_2 \mapsto a_2e^{\delta_A}, \quad a_3 \mapsto a_3e^{\delta_A}.
\ee
\item Scaling of $U$:
\be
U \mapsto U + \delta_U, \quad B \mapsto Be^{\delta_U}, \quad a_2 \mapsto a_2e^{\delta_U}.
\ee
\item Scaling of $W$:
\be
W \mapsto W + \delta_W, \quad a_3 \mapsto a_3e^{\delta_W}.
\ee
\item Scaling of $f$:
\be
f \mapsto \frac{f}{\delta_f^2}, \quad r \mapsto \frac{r}{\delta_f}.
\ee
\end{itemize}
The parameters $\delta_f$, $\delta_A$, $\delta_U$, $\delta_W$ impose the appropriate boundary conditions.
\subsubsection{Horizon boundary conditions}
Imposing regularity of the solutions at the horizon $r=r_h$, and assuming $A_h = 0$, $\dot f_h = 1$, $W_h = 0$ and $U_h = 0$ before rescaling, the following boundary conditions can be obtained:
\be
\dot A_h = \frac{V_g(\lambda_h)}{3} - \frac{x_fV_f(\lambda_h,\tau_h)\left(3Q_h^2 - 1\right)}{6Q_h},
\ee
\be
\dot U_h = -\frac{a_2^2Z(\lambda_h)}{2},
\ee
\be
\dot W_h = \frac{x_fV_f(\lambda_h,\tau_h)\left(Q_h^2 - 1\right)}{2Q_h} - \frac{a_3^2Z(\lambda_h)}{2},
\ee
\begin{align}
\dot\lambda_h & = \frac{3\lambda_h^2}{8}\left[-\partial_\lambda V_g(\lambda_h) + \frac{Y_h\partial_\lambda Z(\lambda_h)}{2}\right.\nonumber \\
& \quad + \left.x_fV_f(\lambda_h,\tau_h)\left(Q_h\partial_\lambda\log V_f(\lambda_h,\tau_h) + \frac{Q_h^2-1 }{Q_h}\partial_\lambda\log w(\lambda_h)\right)\right],
\end{align}
\be
\dot\tau_h = \frac{\partial_\tau\log V_f(\lambda_h,\tau_h)}{\kappa(\lambda_h)}.
\ee
These leave as free parameters to choose $\lambda_h$, $\tau_h$ and $B$, as well as either $a_2$ or $a_3$\@.
\subsubsection{Extraction of quantities}
Temperature, entropy, quark mass and chiral condensate can be extracted the usual way:\footnote{Throughout this section, tildes denote quantities which are \emph{not} rescaled, enabling us to express everything in terms of the nonrescaled variables.}
\be
\frac{T}{\Lambda} = \frac{\dot f_h}{4\pi} = \frac{e^{\delta_A}}{4\pi\delta_f}\frac{\mathrm{d}\tilde f}{\mathrm{d}\tilde r} = \frac{e^{\delta_A}}{4\pi\delta_f}.
\ee
\be
\frac{S}{4\pi M^3N_c^2\Lambda^3} = \exp(3A_h + U_h + W_h) = \exp(3\tilde A_h + 3\delta_A + \tilde U_h + \delta_U + \tilde W_h + \delta_W) = \exp(3\delta_A + \delta_U + \delta_W),
\ee
\be
\frac{m_q}{\Lambda} = \lim_{A \rightarrow \infty}\tau\mathcal{L}_\text{UV}^{-2}e^{\tilde A + \delta_A}\left(\tilde A + \delta_A - \log(\mathcal{L}_\text{UV}\Lambda)\right)^{\gamma_0/b_0}.
\ee
\begin{align}
\frac{\langle\bar qq\rangle}{\Lambda^3} & = \lim_{A\rightarrow\infty}\frac{\exp(2A)\left(A - \log(\mathcal{L}_\text{UV}\Lambda)\right)^{-\gamma_0/b_0}}{2\mathcal{L}_\text{UV}^3\left(\frac{\gamma_0}{b_0} - A + \log(\mathcal{L}_\text{UV}\Lambda)\right)} \\
& \quad \times \left[\exp(A)\left(\frac{\gamma_0}{b_0} + A - \log(\mathcal{L}_\text{UV}\Lambda)\right)\tau + \mathcal{L}_\text{UV}(-A + \log(\mathcal{L}_\text{UV}\Lambda))\tau'\right],
\end{align}
where $b_0$ and $\gamma_0$ are respectively  the first coefficient of the large-N QCD beta function and the leading coefficient of the anomalous dimension of the quark condensate. The magnetic field and the magnetization can be obtained as follows:
\be
\frac{B}{\Lambda^2} = \tilde Be^{2\delta_A + \delta_U},
\ee
\begin{align*}
\frac{4\pi M}{\Lambda^2} & = B\int_{r_h}^{r_b}\mathrm{d}r\frac{e^{A - U + W}x_fV_f(\lambda,\tau)w(\lambda)^2G}{Q}, \\
& = e^{2\delta_A + \delta_U}\tilde B\frac{e^{-\delta_U + \delta_W}}{\delta_f}\int_{r_h}^{r_b}\mathrm{d}\tilde r\frac{e^{\tilde A - \tilde U + \tilde W}x_fV_f(\lambda,\tau)w(\lambda)^2G}{Q}, \\
& = e^{2\delta_A + \delta_U}\tilde B\frac{e^{-\delta_U + \delta_W}}{\delta_f}\int_{A_h}^{A_b}\mathrm{d}\tilde A\frac{\tilde qe^{-\tilde U + \tilde W}x_fV_f(\lambda,\tau)w(\lambda)^2G}{Q}.
\end{align*}
The axion is given by
\be
a_2 = e^{\delta_A + \delta_U}\tilde a_2,
\ee
\be
a_3 = e^{\delta_A + \delta_W}\tilde a_3.
\ee
The horizon value of the derivative of $A_S = A + \frac{2}{3}\log\lambda$, which is important for confinement, is given by
\be
\frac{\mathrm{d}A_{S,h}}{\mathrm{d}r} = \frac{\mathrm{d}\tilde r}{\mathrm{d}r}\frac{\mathrm{d}A_{S,h}}{\mathrm{d}\tilde r} = \delta_fe^{\delta_A}\frac{\mathrm{d}A_{S,h}}{\mathrm{d}\tilde r}.
\ee

\subsection{Choice of potentials} \label{app:potentials}

The potentials used in this article are given by~ \cite{Alho:2012mh,Gursoy:2012bt,Alho:2013hsa,Gursoy:2016ofp}
\begin{eqnarray}
\label{Vf0SB}
V_g(\lambda)&=&{12\over \mathcal{L}_0^2}\biggl[1+{88\lambda\over27}+{4619\lambda^2
\over 729}{\sqrt{1+\ln(1+\lambda)}\over(1+\lambda)^{2/3}}\biggr]\, , \\
 V_{f}(\lambda,\tau)& =& {12\over x_f \mathcal{L}_{UV}^2}\biggl[{\mathcal{L}_{UV}^2\over\mathcal{L}_0^2}
-1+{8\over27}\biggl(11{\mathcal{L}_{UV}^2\over\mathcal{L}_0^2}-11+2x_f \biggr)\lambda\nn\\
 &&+{1\over729}\biggl(4619{\mathcal{L}_{UV}^2\over \mathcal{L}_0^2}-4619+1714x_f - 92x_f^2\biggr)\lambda^2\biggr]\,e^{-a_0\tau^2} \, , \\
 \kappa(\l) &=& {[1+\ln(1+\l)]^{-1/2}\over[1+\frac{3}{4}(\frac{115-16x_f }{27}-{1\over 2})\l]^{4/3}}\, ,\label{kappaa}\\
 w(\l) &=& \kappa(c\, \l)\,,\label{wpotential}\\
Z(\lambda)&=&1+\frac{\lambda ^4}{10}\,, \label{Zdef}
 \end{eqnarray}
where
\be
 a_0 = \frac{3}{2\mathcal{L}_{UV}^2}\, ,\qquad\mathcal{L}_{UV}^3 = \mathcal{L}_0^3 \left( 1+{7 x_f \over 4} \right) \, .
\label{adsrad}
\ee
The parameter $\mathcal{L}_0$ is the AdS radius for $x_f=0$, which we set to one in our numerics. In a similar fashion, $\mathcal{L}_{UV}$ is the AdS radius for $x_f\neq0$\@. Notice that we have set the overall constant in $Z(\lambda)$ ($Z_0$ in the notation of \cite{Gursoy:2012bt}) to unity, since it can be reabsorbed in the normalization of $\chi$.

\section{Details of the anisotropic observables}

\subsection{Holographic thermodynamics} \label{app:thermodynamics}

In this Appendix we discuss how the thermodynamics in the presence of the anisotropic parameter and the magnetic field is computed in the holographic model. This analysis should be compared to the isotropic analysis in~\cite{Gursoy:2008za}\@.

We start with the case where $a= a_\parallel$ is turned on, such that $\chi = a x_3$ so that the magnetic field and  the axion field generate an anisotropy in the same spatial direction. We also need to take into account the Gibbons-Hawking term, which in this case becomes
\be
 S_{GH} = -2 M^3N_c^2 \int d^4 x \sqrt{-\det h} K = -M^3N_c^2 \beta V_3 e^{3 A(\epsilon)+W(\epsilon)}(8 f(\epsilon)A'(\epsilon)+2 f(\epsilon)W'(\epsilon) + f'(\epsilon)) \,,
\ee
where $V_3$ is the volume of the space.
Evaluating the on-shell action using equations of motion, we find for the (nonrenormalized) free energy density
\begin{align}
 F &= -\beta^{-1}V_3^{-1}\left(S_g+S_f+S_{GH}\right)\big|_\mathrm{on-shell} &\\
 \label{Omega1}
&=  M^3N_c^2e^{3 A(\epsilon)+W(\epsilon)} \left(6 f(\epsilon) A'(\epsilon)+f'(\epsilon)\right)
 -a M_a&\\
 &=M^3N_c^2e^{3 A(\epsilon)+W(\epsilon)} \left(6 f(\epsilon) A'(\epsilon)+ 2f(\epsilon) W'(\epsilon)\right) -s T & \\
 &=M^3N_c^2e^{3 A(\epsilon)+W(\epsilon)} \left(6 f(\epsilon) A'(\epsilon)+ 2f(\epsilon) W'(\epsilon)+f'(\epsilon)\right) - B M_B
\label{Omega3}
\end{align}
where $\epsilon$ is a UV cutoff.
Here $M_B$ is the magnetization and $M_a$ is its analogue for the anisotropic parameter, defined in~\eqref{Madef} and~\eqref{MBdef}, respectively. Notice that the free energy can be expressed solely in terms of the boundary geometry if $a=0$, $T=0$, or $B=0$, but not in general. Moreover those terms in the various expressions, which are not boundary data, correspond to Legendre transformations. The equivalence of the expressions \eqref{Omega1}--\eqref{Omega3} reflect the equations
\begin{align}
\label{pareq1}
 \frac{d}{dr}\left[e^{3 A(r)+W(r)}(f'(r)-2f(r)W'(r))\right]
 &= a^2e^{3 A(r)-W(r)} Z(\lambda (r))&\\
  \frac{d}{dr}\left[e^{3 A(r)+W(r)}f'(r)\right]
  &= B^2 \frac{ x_fG(r) e^{A(r)+W(r)} w(\lambda (r))^2 V_f(\lambda (r),\tau (r))}{Q(r)}
\label{pareq2}
\end{align}
which follows from the Einstein equations. One of them can be immediately integrated once if either $a=0$ or $B=0$.

We then discuss the case $a= a_\perp$, where $\chi = ax_2$ so that (assuming $B \ne 0$) all rotation symmetries are broken.
In this case we find for the Gibbons-Hawking term
\be
S_{GH} =  -M^3N_c^2 \beta V_3 e^{3 A(\epsilon)+U(r)+W(\epsilon)}(8 f(\epsilon)A'(\epsilon)+2 f(\epsilon)U'(\epsilon)+2 f(\epsilon)W'(\epsilon) + f'(\epsilon)) \ .
\ee
The free energy density can again be written in various forms, for example
\begin{align}
 F &= M^3N_c^2e^{3 A(\epsilon)+U(\epsilon)+W(\epsilon)} \left(6 f(\epsilon) A'(\epsilon)+ 2f(\epsilon) U'(\epsilon)+f'(\epsilon)\right)  &\\
 &=M^3N_c^2e^{3 A(\epsilon)+U(\epsilon)+W(\epsilon)} \left(6 f(\epsilon) A'(\epsilon)+ 2f(\epsilon) U'(\epsilon)+ 2f(\epsilon) W'(\epsilon)\right)  -s T & \\
  &= M^3N_c^2e^{3 A(\epsilon)+W(\epsilon)} \left(6 f(\epsilon) A'(\epsilon)+f'(\epsilon)\right)
 -a M_a&\\
 &=M^3N_c^2e^{3 A(\epsilon)+W(\epsilon)} \left(6 f(\epsilon) A'(\epsilon)+ 2f(\epsilon) U'(\epsilon)+ 2f(\epsilon) W'(\epsilon)+f'(\epsilon)\right) - B M_B
\end{align}
by using the identities
\begin{align}
\label{pereq1}
 \frac{d}{dr}\!\left[e^{3 A(r)+U(r)+W(r)}(f'(r)-2f(r)W'(r))\right] &= 0&\\
  \frac{d}{dr}\!\left[e^{3 A(r)+U(r)+W(r)}f(r)W'(r)\right] &=
  B^2 \frac{ x_f G(r) e^{A(r)+W(r)-U(r)} w(\lambda (r))^2 V_f(\lambda (r),\tau (r))}{ 2 Q(r)} &\\
    \frac{d}{dr}\!\left[e^{3 A(r)+U(r)+W(r)}f(r)U'(r)\right] &=-\frac{1}{2} a^2 e^{3 A(r)+W(r)-U(r)} Z(\lambda (r)) \,.&
\label{pereq3}
\end{align}

We continue by discussing the holographic renormalization, which we carry out by subtracting a reference background, following~\cite{Gursoy:2008za}. The renormalization can be worked out independently of the direction of $a$ and $B$ since the relevant potential, i.e., the energy density
takes the same form for both cases defined above. First we recall that the UV asymptotics of the background are given by
\begin{align}
 b_0\lambda &= -\frac{1}{ \log r \Lambda} +\mathcal{O}\left(\frac{1}{(\log r \Lambda)^2}\right)  -\frac{45}{8}\mathcal{G}\frac{r^4}{\ell^3} \left(1+\mathcal{O}\left(\frac{1}{\log r \Lambda}\right)\right) &\\
 e^{A} &= \frac{\ell}{r}\left[1 + \frac{4}{9\log r\Lambda} +\mathcal{O}\left(\frac{1}{(\log r \Lambda)^2}\right) +\mathcal{G} \frac{r^4}{\ell^3} \left(1+\mathcal{O}\left(\frac{1}{\log r \Lambda}\right)\right)\right] &\\
   f & = 1+\mathcal{O}(B^2 r^4 \log r)- \frac{C_f}{4} \frac{r^4}{\ell^3} \left(1+\mathcal{O}\left(\frac{1}{\log r \Lambda}\right)\right) &\\
  U & = \mathrm{source\ term} + \frac{C_U}{4} \frac{r^4}{\ell^3} \left(1+\mathcal{O}\left(\frac{1}{\log r \Lambda}\right)\right) &\\
  W & = \mathrm{source\ term} + \frac{C_W}{4} \frac{r^4}{\ell^3} \left(1+\mathcal{O}\left(\frac{1}{\log r \Lambda}\right)\right)\,, &
\end{align}
where we omitted the $a$ and $B$ dependent source terms which we will not need. The reference background needs to be subtracted at the UV cutoff with the same values for the scalar fields $\lambda$ and $\tau$, keeping also the four-volume fixed. We restrict here to zero quark mass, so $\tau$ will not enter the UV asymptotics and can be neglected (see~\cite{Jarvinen:2015ofa} for the discussion of the dependence on quark mass). Taking into account the renormalization conditions\footnote{The reference values of $a$ and $B$ also need to be chosen correctly in order to ensure cancellation of UV divergences. In the case of $a$ this is a bit subtle because apart from the $\mathcal{O}\left(a^2\right)$ term also the $\mathcal{O}\left(a^4\right)$ term in the UV expansions creates a divergence, see~\cite{Gursoy:2018ydr}. This subtlety does not affect directly the analysis here.} for the reference volume and temperature~\cite{Gursoy:2008za}, we obtain
\begin{align}
 \mathcal{E}_\mathrm{ren}& = M^3N_c^2e^{3 A(\epsilon)+U(\epsilon)+W(\epsilon)} \left(6 f(\epsilon) A'(\epsilon)+ 2f(\epsilon) U'(\epsilon)+ 2f(\epsilon) W'(\epsilon)\right) &\\
 & - M^3N_c^2e^{4 A(\epsilon)-\tilde A(\tilde \epsilon)+U(\epsilon)+W(\epsilon)}\sqrt{\frac{f(\epsilon)}{\tilde f(\tilde \epsilon)}} \left(6 \tilde f(\tilde\epsilon) \tilde A'(\tilde\epsilon)+ 2\tilde f(\tilde\epsilon) \tilde U'(\tilde \epsilon)+ 2\tilde f(\tilde\epsilon) \tilde W'(\tilde \epsilon)\right)
\end{align}
where the fields with tildes are those of the reference background and the cutoff $\tilde \epsilon$ is determined by the condition $\lambda(\epsilon) = \tilde \lambda (\tilde \epsilon)$, which gives
\be
 \frac{\tilde \epsilon}{\epsilon} = 1 -\frac{45}{8}\mathcal{G}\frac{\epsilon^4}{\ell^3} (-\log \epsilon \Lambda)^2\left(1+\mathcal{O}\left(\frac{1}{\log \epsilon \Lambda}\right)\right) \,.
\ee
Inserting the UV expansions and taking $\epsilon \to 0$ gives
\be
  \frac{\mathcal{E}_\mathrm{ren}}{M^3N_c^2} = \frac{3}{4} \left(C_f-\tilde C_f\right)+15 \left(\mathcal{G}- \mathcal{\tilde G}\right)
  + 2\left(C_U-\tilde C_U\right)  + 2\left(C_W-\tilde C_W\right)
\ee
where the tilded coefficients are those of the reference background. Choosing the zero value of the energy density to match with the reference, we therefore define the final energy density as
\be
 \overline{\mathcal{E}} = M^3N_c^2\left(\frac{3}{4}C_f+15\mathcal{G}+2C_U+2C_W \right)\,.
\ee
The renormalized free energy density is then
\be
 \overline{F} =  \overline{\mathcal{E}} - sT  \,.
\ee

We proceed by computing the complete stress-energy tensor by using the dictionary, i.e.,
\be
 T_{\mu\nu} = \frac{2}{V_4} \frac{\delta S_\mathrm{ren}}{\delta \gamma^{(0)}_{\mu\nu}}
\ee
where $V_4$ is the four-volume and $\gamma^{(0)}_{\mu\nu}$ are the source terms for the spatial components of the metric. To do this explicitly, we may first start by reparameterizing the metric as
\be
 ds^2 = e^{2\bar A(r)}\left(dr^2 -f_t(r)dt^2+\sum_{k=1}^3 f_k(r) dx_k^2 \right)
\ee
where each of the $f_i$ has a constant source term; the variations to compute the stress-energy tensor are taken with respect to them. All other sources, including the temperature, are to be held fixed when varying. After obtaining the result, we may revert back to the original definitions for the metric. As the dust clears we find that (with the normalization convention that the  contributions from the reference background are dropped
\begin{align}
 T_{00} &= \overline{\mathcal{E}} =  M^3N_c^2\left(\frac{3}{4}C_f+15\mathcal{G}+2C_U+2C_W \right) & \\
 T_{11} &= p_1 = M^3N_c^2\left(\frac{1}{4}C_f-15\mathcal{G}-2C_U-2C_W \right) & \\
 T_{22} &= p_2 = M^3N_c^2\left(\frac{1}{4}C_f-15\mathcal{G}-2C_W \right) &  \\
 T_{33} &= p_3 = M^3N_c^2\left(\frac{1}{4}C_f-15\mathcal{G}-2C_U \right) \,. &
\end{align}
The pressure anisotropy is, as expected, given in terms of the VEVs for $U$ and $W$:
\be
 p_2-p_1 = 2 M^3N_c^2 C_U \,,\qquad p_3-p_1 = 2 M^3N_c^2 C_W \,.
\ee

Let us then write down the results which depend on the orientation of the asymmetry. For the first case (parallel, $a=a_\parallel$), we have $C_U=0$ and the equations~\eqref{pareq1} and~\eqref{pareq2} imply
\be
 M^3N_c^2C_f = s T
 - B \overline{M}_B \,,\qquad 2 M^3N_c^2 C_W = B \overline{M}_B - a \overline{M}_a
\ee
where $\overline{M}_B$ and $\overline{M}_a$ are the renormalized quantities with respect to the reference background. Denoting $p_1 = p_2 =p_\perp$, $p_3=p_\parallel$, we find that
\be
 \overline{\mathcal{E}} + p_\perp = sT
 -B \overline{M}_B \,,\qquad \overline{\mathcal{E}} + p_\parallel = sT
 -a \overline{M}_a
\ee
or in terms of the free energy density and pressure
\be
 \overline{F} = -  p_\perp -B \overline{M}_B =-p_\parallel-a \overline{M}_a \,,\qquad p_\parallel- p_\perp = B \overline{M}_B - a \overline{M}_a \,.
\ee

For the second case (perpendicular, $a=a_\perp$) equations~\eqref{pereq1}--\eqref{pereq3} imply
\be
M^3N_c^2C_f = s T
- B \overline{M}_B \,,\qquad 2 M^3N_c^2 C_U =  - a \overline{M}_a\,,\qquad 2 M^3N_c^2 C_W = B \overline{M}_B
\ee
so that
\be
  \overline{\mathcal{E}} + p_1 = sT
  -B \overline{M}_B \,,\qquad \overline{\mathcal{E}} + p_2 = sT
  -B \overline{M}_B-a \overline{M}_a\,,\qquad \overline{\mathcal{E}} + p_3 = sT
\ee
or equivalently
\be
 \overline{F} = -  p_1 -B \overline{M}_B =-p_2-B \overline{M}_B-a \overline{M}_a = -p_3 \,.
\ee
The pressure differences satisfy
\be
  p_3-p_1 = B \overline{M}_B \,,\qquad  p_2-p_1 = -a \overline{M}_a \,.
\ee
When $a=0$ ($B=0$), our results are consistent with those of~\cite{Ammon:2012qs} (\cite{Mateos:2011tv}), respectively.

\subsection{Shear viscosity} \label{app:eta}

In this Appendix we sketch the computation of the shear viscosity in anisotropic backgrounds. As it turns out, it is necessary to discuss separately the cases where the anisotropy due to the bulk axion field is perpendicular ($a_\perp$) and parallel ($a_\parallel$).

We start with the perpendicular case, where, somewhat counterintuitively, the results can be written in a simpler form. We write the fluctuations as
\be
 \delta g_{12} = e^{2A(r)}e^{-i\omega t} \zeta_{12}(r) \,, \quad \delta g_{13} = e^{2A(r)+2W(r)}e^{-i\omega t} \zeta_{13}(r)  \,, \quad \delta g_{23} = e^{2A(r)+2W(r)}e^{-i\omega t} \zeta_{23}(r) \,.
\ee
Then the fluctuation equations may be written as follows:
\begin{align}
 \frac{d}{dr}\left[f(r)e^{3A(r)+W(r)-U(r)}\zeta_{12}'(r)\right] + \omega^2\, \frac{e^{3A(r)+W(r)-U(r)}}{f(r)}\zeta_{12}(r) &= 0 \,,&\\
 \frac{d}{dr}\left[f(r)e^{3A(r)+3W(r)+U(r)}\zeta_{13}'(r)\right] + \omega^2\, \frac{e^{3A(r)+3W(r)+U(r)}}{f(r)}\zeta_{13}(r) &= 0 \,,&\\
 \frac{d}{dr}\left[f(r)e^{3A(r)+3W(r)-U(r)}\zeta_{23}'(r)\right] + \omega^2\, \frac{e^{3A(r)+3W(r)-U(r)}}{f(r)}\zeta_{23}(r) &= 0 \,.&
\end{align}
where we used the Einstein equations~\eqref{Ueq} and~\eqref{Weq} to simplify the result. We see that the only difference with respect to the isotropic case is a slight modification of the warp factors appearing in these equations. The rest of the computation proceeds as in the isotropic case, and recalling that $s = 4\pi M^3N_c^2e^{3A(r_h)+W(r_h)+U(r_h)}$, we find the formulas~\eqref{etaxy}--\eqref{etayz}. Notice that these formulas agree with the literature for the case when either $a=0$ or $B=0$~\cite{Rebhan:2011vd,Jain:2015txa}.

We then discuss the parallel case, $a=a_\parallel$. Now $U(r)=1$ and by symmetry $\eta_{13}=\eta_{23}$ so it is enough to study the following fluctuations:
\be
\delta g_{12} = e^{2A(r)}e^{-i\omega t} \zeta_{12}(r) \,, \quad \delta g_{13} = e^{2A(r)+2W(r)}e^{-i\omega t} \zeta_{13}(r) = e^{2A(r)}e^{-i\omega t}\bar \zeta_{13}(r)\,.
\ee
The first fluctuation equation reads simply
\be
  \frac{d}{dr}\left[f(r)e^{3A(r)+W(r)}\zeta_{12}'(r)\right] + \omega^2\, \frac{e^{3A(r)+W(r)}}{f(r)}\zeta_{12}(r) = 0
\ee
and analyzing this leads to the standard result,
\be
 \frac{\eta_{12}}{s} = \frac{1}{4\pi}\,.
\ee
For the second fluctuation equation we find two equivalently complicated forms:
\begin{align}
\label{zetaeq}
  &\frac{d}{dr}\left[f(r)e^{3A(r)+3W(r)}\zeta_{13}'(r)\right] + \omega^2\, \frac{e^{3A(r)+3W(r)}}{f(r)}\zeta_{13}(r) = a^2 e^{3A(r)+W(r)} Z(\lambda(r)) \zeta_{13}(r)\,, & \\
& \frac{d}{dr}\left[f(r)e^{3A(r)-W(r)}\bar\zeta_{13}'(r)\right] + \omega^2\, \frac{e^{3A(r)-W(r)}}{f(r)}\bar\zeta_{13}(r) &\nonumber\\
&\qquad\qquad\qquad\qquad\qquad\qquad = B^2\frac{x_f e^{A(r)-W(r)}V_f(\lambda(r),\tau(r))w(\lambda(r))^2G(r)}{Q(r)}  \bar\zeta_{13}(r)\,. &
\label{zetabareq}
\end{align}
Notice that, unlike in the perpendicular case above, the only relevant Einstein equation~\eqref{Weq} now depends explicitly on both $a$ and $B$ and is therefore not useful to simplify the fluctuation equations further. Following~\cite{Gubser:2008sz} we find the following formulas
\be \label{etaparallel}
 \frac{\eta_{13}}{s} = \frac{1}{4\pi}e^{2 W(r_h)}c_a^2 = \frac{1}{4\pi}e^{-2 W(r_h)}c_B^2 \,.
\ee
Here the real coefficients $c_{a,B}$ are defined as
\be
 c_a = \frac{\zeta_{13}^{(0)}(r_h)}{\zeta_{13}^{(0)}(0)} \,,\qquad  c_B = \frac{\bar\zeta_{13}^{(0)}(r_h)}{\bar\zeta_{13}^{(0)}(0)} \,,
\ee
where $\zeta_{13}^{(0)}(r)$ and  $\bar\zeta_{13}^{(0)}(r)$ are the IR regular solutions to~\eqref{zetaeq} and~\eqref{zetabareq}, respectively, at $\omega = 0$. Notice that when $a=0$ ($B=0$) we have that $c_a=1$ ($c_B=1$) respectively, because the regular solution to the relevant fluctuation equation is constant. This ensures that the result~\eqref{etaparallel} is consistent with~\eqref{etaxy}--\eqref{etayz}.

\subsection{Butterfly velocity \label{app:vb}}

In this Appendix we derive a formula for the components of the butterfly velocity $v_B^i$ for general anisotropic backgrounds, using ideas of subregion duality \cite{Mezei:2016wfz}.

We start with a generic $(d+1)-$dimensional black brane metric
\be\label{ansatzgeneral}
ds^2=-g_{tt}(r)dt^2+g_{rr}(r)dr^2+\sum_{i=1}^{d-1} g_{ii}(r) dx_i^2\,,
\ee
where $d$ is the number of dimensions in the dual CFT (including time). We assume that in this coordinate system the boundary (UV) is located at $r=0$, and the horizon (IR) is located at $r=r_h$. For our particular case, we have that $d=4$ and
\be
g_{tt}(r)=e^{2A(r)}f(r)\,, \qquad g_{rr}(r)=\frac{e^{2A(r)}}{f(r)}\,,\qquad g_{ii}(r)=e^{2A(r)+2X_i(r)}\,,
\ee
where $X_i(r)=\{0,U(r),W(r)\}$. However, we will do these replacements only at the end of the calculation. For now we will continue assuming a metric of the form (\ref{ansatzgeneral}) with the intention of obtaining a more generic formula for $v_B^i$.

Following \cite{Mezei:2016wfz}, we now specialize to the near-horizon geometry, where
\be
g_{tt}(r)\simeq c_0(r_h-r)\,,\quad g_{rr}(r)\simeq \frac{c_1}{r_h-r}\,,\quad g_{ii}(r)\simeq g_{ii}(r_h)-g_{ii}'(r_h)(r_h-r)\,,
\ee
so that
\be
ds^2\simeq -c_0(r_h-r)dt^2+\frac{c_1}{r_h-r}dr^2+\sum_{i=1}^{d-1} \left[g_{ii}(r_h)-g_{ii}'(r_h)(r_h-r)\right]dx_i^2\,.
\ee
Here $c_0$ and $c_1$ are two positive constants, in terms of which the inverse Hawking temperature reads
\be
\beta=4\pi\sqrt{\frac{c_1}{c_0}}\,.
\ee
Now, we define a Rindler coordinate $\rho$ according to
\be
(r_h-r)=\left(\frac{2\pi}{\beta}\right)^2\frac{\rho^2}{c_0}
\ee
so that the near-horizon metric becomes
\be
ds^2\simeq -\left(\frac{2\pi}{\beta}\right)^2\rho^2dt^2+d\rho^2+\sum_{i=1}^{d-1} \left[g_{ii}(r_h)+\frac{g_{ii}'(r_h)}{g_{tt}'(r_h)}\left(\frac{2\pi}{\beta}\right)^2\rho^2\right]dx_i^2\,.
\ee
In the last term we have identified $c_0=-g_{tt}'(r_h)$.

Next, we consider an infalling particle that arises as a result of a local perturbation. The particle gets blue shifted and
approaches the horizon exponentially
\be
\rho(t)=\rho_0 e^{-\frac{2\pi}{\beta}t}\,.
\ee
The butterfly velocity $v_B$ can then be computed by finding the smallest entanglement wedge that contains the particle at late times \cite{Mezei:2016wfz}. We parametrize the embedding function as $\rho(x^i)$, and pick local coordinates $\xi^i=x^i$. The area functional in the near-horizon region (small $\rho$) then becomes
\be
\text{Area}(\gamma_A) = \sqrt{\det{g_{ij}(r_h)}} \int d^{d-1}x \left[1+\frac{1}{2}\left( \frac{2 \pi}{\beta} \right)^2 \frac{\rho^2}{g_{tt}'(r_h)}\sum_{i=1}^{d-1}\frac{g_{ii}'(r_h)}{g_{ii}(r_h)}+\frac{1}{2}\sum_{i=1}^{d-1}\frac{(\partial_i \rho)^2}{g_{ii}(r_h)} \right].
\ee
Notice that since the metric (\ref{ansatzgeneral}) is diagonal, we can replace several factors in the denominators using the inverse metric, e.g., $1/g_{ii}(r)=g^{ii}(r)$.
From this functional we can derive the following equation for the embedding function:
\be
\sum_{i=1}^{d-1}g^{ii}(r_h)\partial_i^2 \rho(x^i) = \mu^2 \rho(x^i)\,,\qquad \mu^2\equiv  \frac{\left(2 \pi/\beta \right)^2}{g_{tt}'(r_h)}\sum_{i=1}^{d-1}g^{ii}(r_h) g_{ii}'(r_h)\,.
\ee
Now, we rescale the coordinates $x^i=\sqrt{g^{ii}(r_h)}\sigma^i$, so that the metric becomes locally isotropic at leading order. The above equation then becomes:
\be
(\nabla_\sigma)^2 \rho = \mu^2 \rho\,,\qquad (\nabla_\sigma)_i=\frac{\partial}{\partial \sigma^i} \,.
\ee
This equation has the following solution:
\be
\rho(\sigma^i)=\rho_\text{min} \frac{\Gamma(n+1)}{2^{-n}\mu^{n}} \frac{I_n(\mu |\sigma|)}{|\sigma|^n}\,,\qquad n\equiv \frac{d-3}{2}\,,
\ee
where $\rho_\text{min}$ is the radius of closest approach to the horizon (located at $\rho=0$) and $I_n$ is a modified Bessel function of the second kind. As argued in \cite{Mezei:2016wfz}, when $\rho\gtrsim\beta$, the surface exits the near-horizon region and approaches the boundary very fast (almost perpendicularly, since an RT surface that probes the near-horizon geometry corresponds to a very large boundary region $A$). We can then estimate the size of the region $A$ in terms of $\rho_\text{min}$ by solving the equation
\be
\beta\simeq \rho_\text{min} \frac{\Gamma(n+1)}{2^{-n}\mu^{n}} \frac{I_n(\mu R_{\sigma})}{R_{\sigma}^n}\,,
\ee
where $R_{\sigma}$ is the size of region $A$ in the coordinates $\sigma^i$. The solution at large $R_{\sigma}$ is:
\be
\rho_\text{min} \simeq e^{-\mu R_{\sigma}}\,.
\ee
Next, we go back to the original coordinates. For anisotropic theories, the size of region $A$ is different along the different directions $x^i$, and so is the associated component of the butterfly velocity $v_B^i$. We denote $R_i$ as the size of the region $A$ along the $x^i$ direction, which is given by $R_i=\sqrt{g^{ii}(r_h)}R_{\sigma}$. Requiring that the infalling particle is contained inside the entanglement wedge, $\rho_\text{min} \leq \rho(t)$, implies
\be
\mu \sqrt{g_{ii}(r_h)}R_i \geq \frac{2\pi}{\beta}t\,,
\ee
which leads to
\be\label{eq:finalvb}
R_i \geq v_{B}^i t\,,\qquad  v_{B}^i\equiv \frac{2\pi/\beta}{\sqrt{g_{ii}(r_h)} \mu}=\sqrt{\frac{g^{ii}(r_h)g_{tt}'(r_h)}{\sum_{k=1}^{d-1}g^{kk}(r_h)g_{kk}'(r_h)}}\,.
\ee
This is the final formula that we were after. Notice that in the isotropic limit it coincides with the formula for the butterfly velocity derived in \cite{Mezei:2016wfz}. It is also consistent with previous results for anisotropic theories, where \emph{only one} direction is anisotropic, including the scaling solutions of \cite{Giataganas:2017koz} and various other models \cite{Jahnke:2017iwi,Avila:2018sqf,Fischler:2018kwt,Inkof:2019gmh}. Finally, specializing to our metric (\ref{ansatz}) we obtain the formulas reported in (\ref{eq:vbour}).

\subsection{Results for $c=0.4$ \label{app:cdepen}}

In section \ref{sec:otherObs} we showed that our results for the string tensions strongly favoured $c=0.25$ after comparing with the lattice data of \cite{Bonati:2014ksa}. However, it is known that our model can exhibit IMC for even slightly higher values of $c$. It is then interesting to explore the effects of this parameter on the behavior of the other observables of interest, and investigate how robust our findings and predictions are. Throughout this appendix we will study in detail this issue for $c=0.4$, which is also known to display both IMC \cite{Gursoy:2016ofp} and IAC \cite{Gursoy:2018ydr}.

\begin{figure}[t!]
\centering
\includegraphics[width=0.85\textwidth]{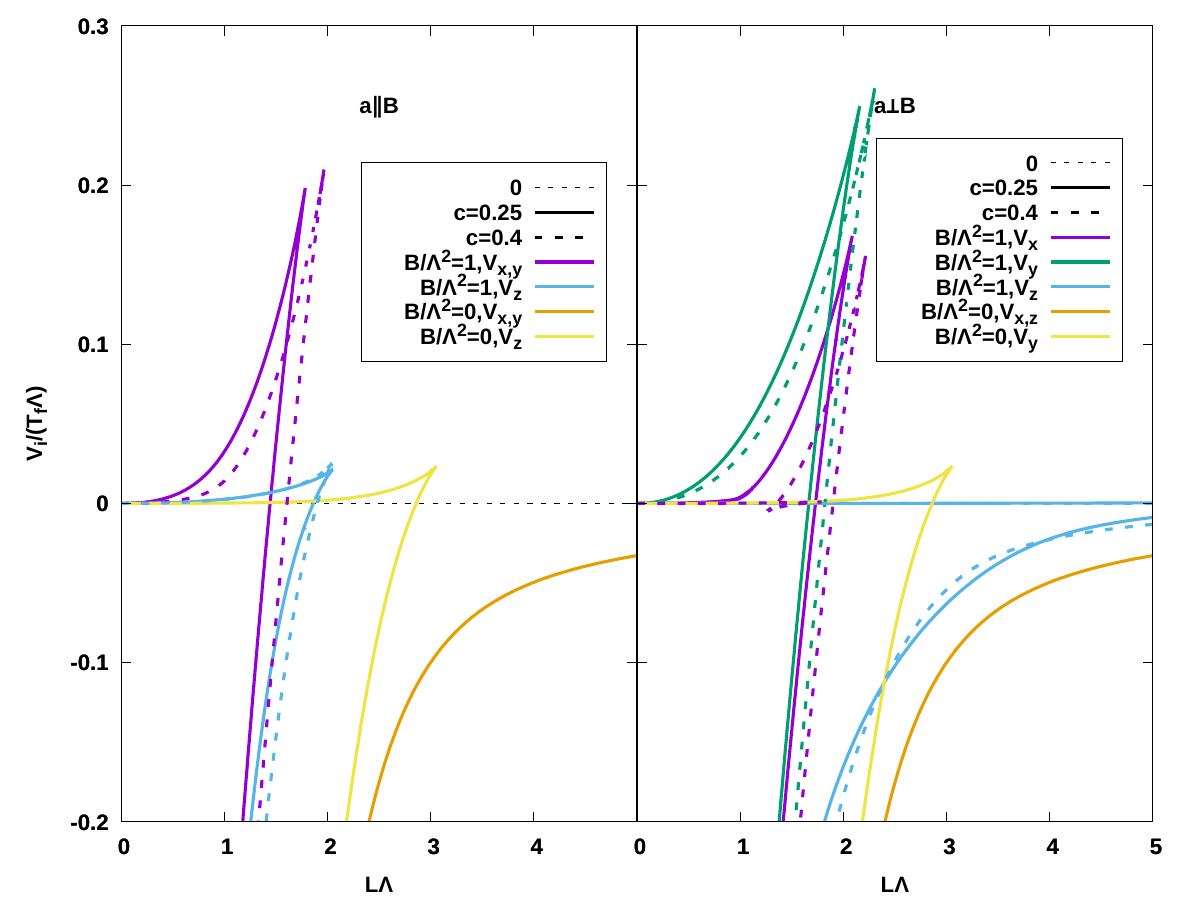}
\caption{Quark-antiquark potential in the presence of anisotropy and magnetic field, for $a=1$ and $B=1$ (in units of $\Lambda\sim1$ GeV). The magnetic field $B$ is aligned along the $z$-direction while the anisotropic deformation is introduced along the $z$- or $y$-direction (parallel and perpendicular configurations, respectively). The solid lines depict results for $c=0.25$, which we have kept for comparison, while dashed lines show the corresponding results for $c=0.4$.
We find the same qualitative features in both cases, perhaps with the only difference that the increase in $c$ pushes the transition to slightly higher values of the separation, making the bound states a bit more stable. \label{fig:potBandaApp}}
\end{figure}

In fig.~\ref{fig:potBandaApp} we plot the results for the quark-antiquark potentials in a situation where both $a$ and $B$ are turned on and can compete with each other. In general, we observe the same qualitative features between the two values of $c$: the tendency to push the transition to shorter distances when both effects are present while increasing the tensions, and the two special cases --- $V_z$ in the case where $a$ and $B$ are perpendicular/parallel to each other --- where $a$ and $B$ act completely constructively/destructively in the IR\@. Perhaps the only distinction that is worth pointing out is the slight increase in the transition distance as we increase $c$, which make the bound states a bit more stable in comparison to the $c=0.25$ case.

\begin{figure}[t!]
\centering
\includegraphics[width=1\textwidth]{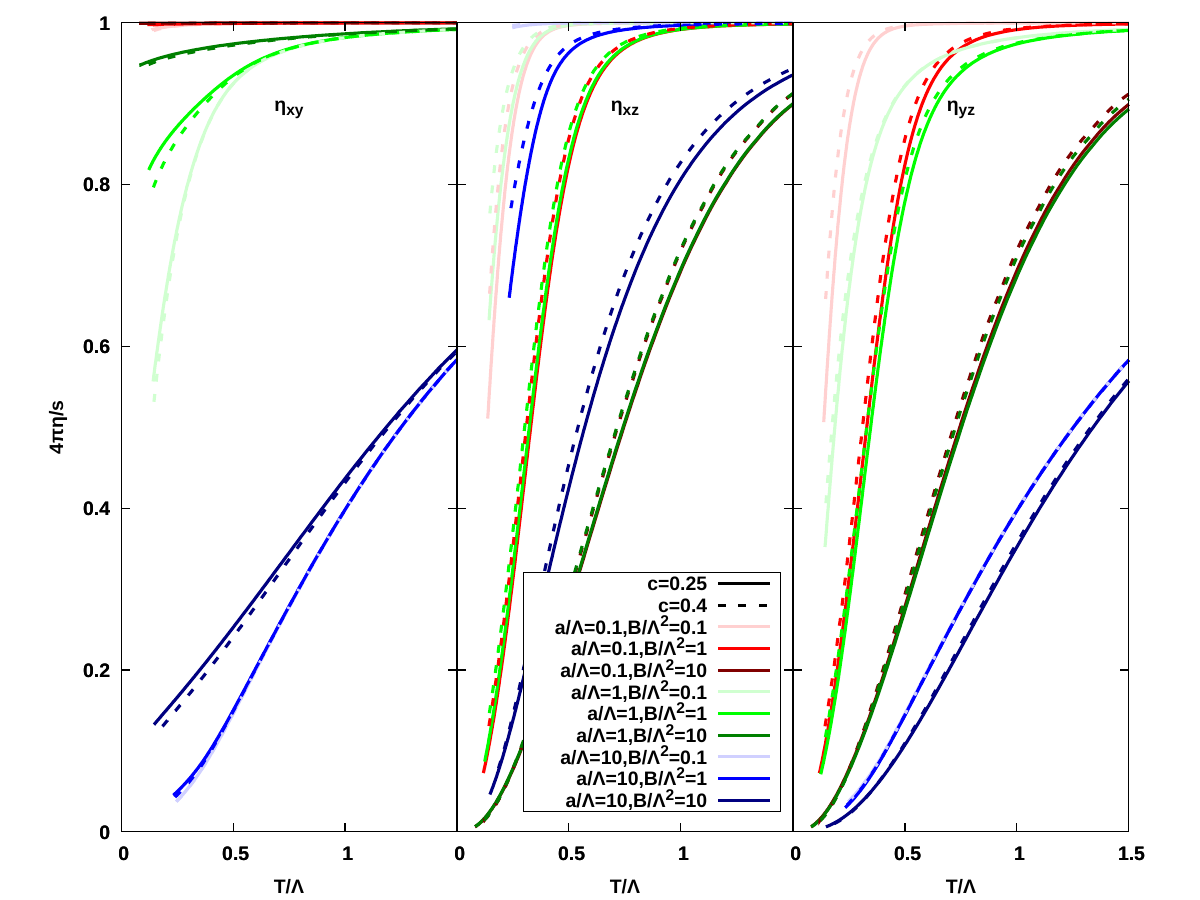}
\caption{Ratio between the components of the shear viscosity tensor and the entropy density for the case where the magnetic field (aligned along the $z$-direction) is perpendicular to the anisotropic deformation (introduced along the $y$-direction). All the plots are given as a function of temperature, in units of $\Lambda\sim1$ GeV, and cut off at the chiral transition temperature so they correspond to the chirally symmetric phase. The solid lines depict results for $c=0.25$, which we have kept for comparison, while dashed lines show the corresponding results for $c=0.4$. We observe all the same qualitative features for the two values of $c$, with some small visual distinction along the flow from the UV to the IR: while $\eta_{xy}$ generally decreases with the increase of $c$, both $\eta_{xz}$ and $\eta_{yz}$ increase at an intermediate regime of energies, while approaching roughly the same constants in the IR. \label{fig:shearApp}}
\end{figure}

\begin{figure}[t!]
\centering
\includegraphics[width=1\textwidth]{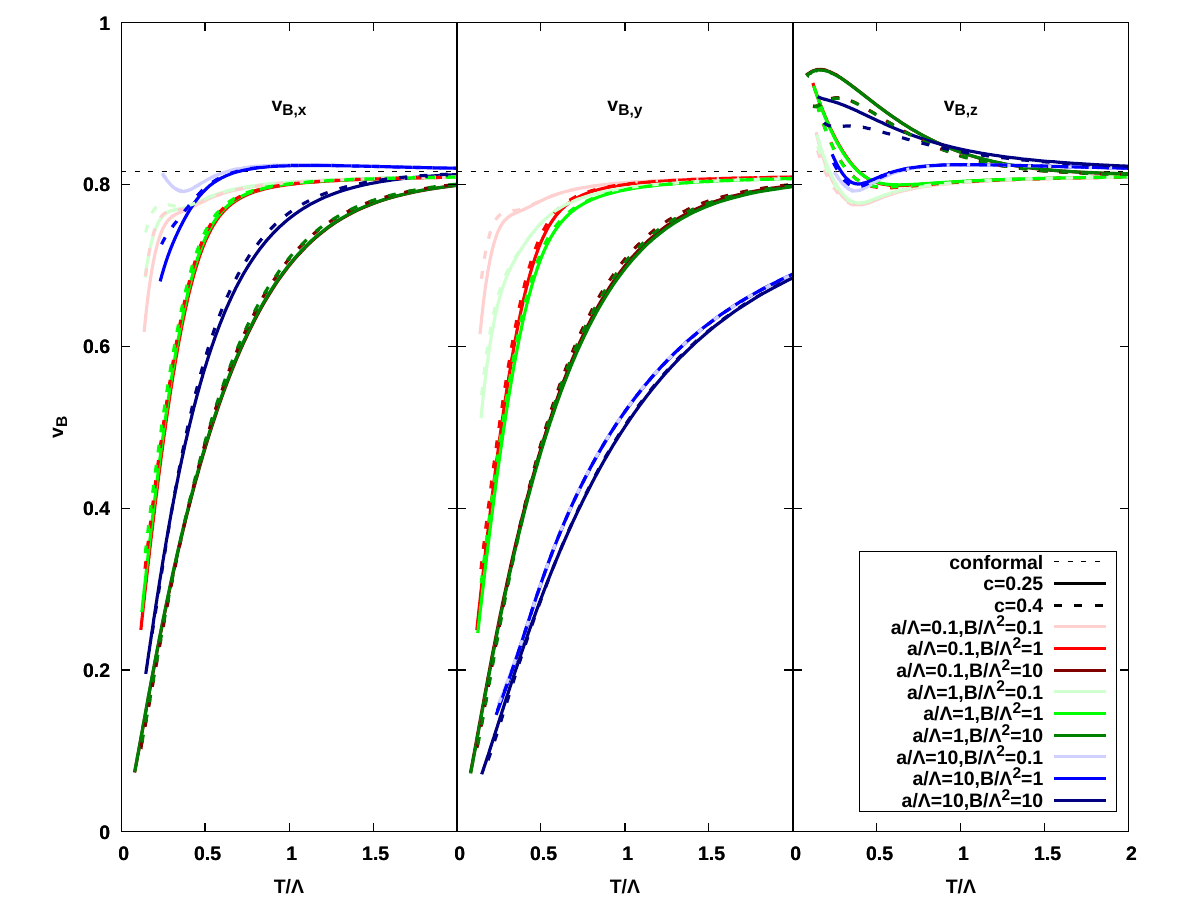}
\caption{Components of the butterfly velocity as a function of temperature (in units of $\Lambda\sim1$ GeV) for the perpendicular configuration. The magnetic field $B$ is aligned along the $z$-direction while the anisotropic deformation is introduced along the $y$-direction. The solid lines depict results for $c=0.25$, which we have kept for comparison, while dashed lines show the corresponding results for $c=0.4$. In both cases we observe that the component $v_B^z$ is enhanced in the IR and exceeds the bound found in \cite{Giataganas:2017koz}, although its maximum value decreases with the increase of $c$. On the other hand, the other components $v_B^x$ and $v_B^y$ show a slight increase with respect to the results for $c=0.25$. \label{fig:bvperpApp}} 
\end{figure}
\begin{figure}[t!]
\centering
\includegraphics[width=0.85\textwidth]{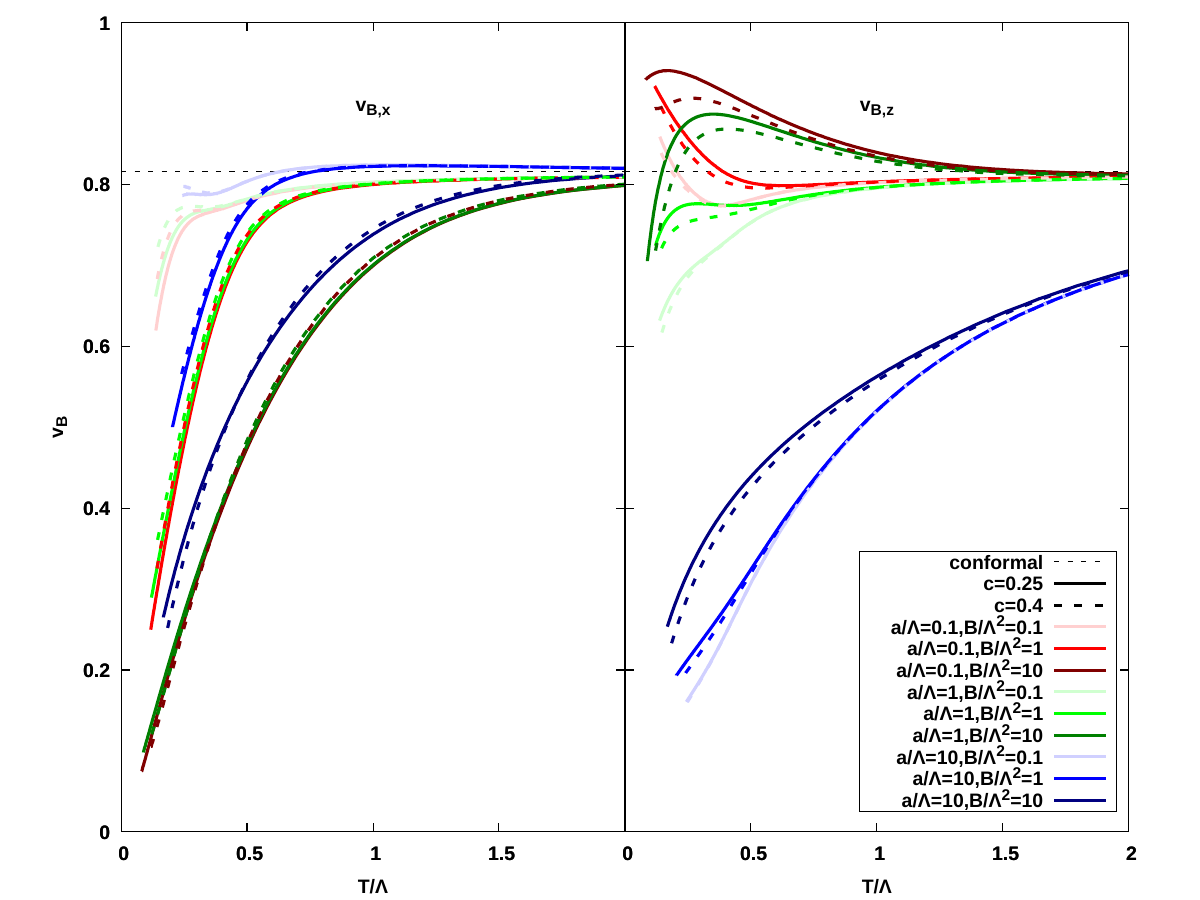}
\caption{Components of the butterfly velocity as a function of temperature (in units of $\Lambda\sim1$ GeV) for the parallel configuration. In this case both the magnetic field $B$ and the anisotropic deformation are introduced along the $z$-direction. The solid lines depict results for $c=0.25$, which we have kept for comparison, while dashed lines show the corresponding results for $c=0.4$. The component $v_B^z$ is also enhanced in the IR and exceeds the bound found in \cite{Giataganas:2017koz}, although it is slightly smaller than in the perpendicular case. In this case, the increase in $c$ also makes $v_B^z$ smaller, while there is no clear pattern for the other component, $v_B^x$.\label{fig:bvparaApp}}
\end{figure}
In fig.~\ref{fig:shearApp} we plot the components of the shear viscosity in the case where the magnetic field is perpendicular to the anisotropic deformation. We observe all the same qualitative features for the two values of $c$: in all cases, the plots exhibit a monotonic flow from $\eta_{ij}/s=1/4\pi$ in the UV to a lower, non-zero value in the IR, before the chiral transition kicks in. Moreover, the three main conclusions reached for $c=0.25$ hold true for $c=0.4$ as well: i) The increase in magnetic field $B$ raises $\eta_{xy}$ and lowers $\eta_{xz}$ and $\eta_{yz}$. ii) The increase in the anisotropic parameter $a$ lowers $\eta_{xy}$ and $\eta_{yz}$ and raises $\eta_{xz}$. And iii) $B$ affects mostly $\eta_{xz}$, while $a$ affects mostly $\eta_{xy}$. Finally, the two effects seem equally strong for $\eta_{yz}$ so there is no clear pattern for this component. The only visual differences between the two values of $c$ are small differences in values along the flow from the UV to the IR: while $\eta_{xy}$ generally decreases with the increase of $c$, both $\eta_{xz}$ and $\eta_{yz}$ increase at an intermediate regime of energies, while approaching roughly the same constants in the IR.

Finally, in figures \ref{fig:bvperpApp} and \ref{fig:bvparaApp} we plot the components of the butterfly velocity in various situations of interest. Both $c=0.25$ and $c=0.4$ show qualitatively similar results; crucially, that the component $v_B^z$ is enhanced in the IR and exceeds the bound found in \cite{Giataganas:2017koz}. However, the maximum value decreases with the increase of $c$, $v_{B}^z\approx 0.907$ for $c=0.4$ while $v_{B}^z\approx 0.942$ for $c=0.25$. In both cases, parallel and perpendicular configurations, the remaining components show a weak dependence on $c$, with a tendency of slightly increasing the values of $v_{B}^x$ and $v_{B}^y$. As explained in the main text, the violation of the bound on the butterfly velocity does not contradict the results of \cite{Giataganas:2017koz} because the IR theory in our case does not belong to the same universality class of those in \cite{Giataganas:2017koz}. We will present a more thorough analysis of these bounds in a more generic class of backgrounds in our upcoming publication \cite{Giataganas:progress}.

\bibliographystyle{ucsd}
\bibliography{refs}

\end{document}